\documentclass[prb,,superscriptaddress,twocolumn]{revtex4-1}
\usepackage{amsmath,amssymb,mathrsfs}
\usepackage{graphicx,import}
\usepackage[usenames,dvipsnames]{color}
\usepackage[colorlinks=true,linkcolor=blue,citecolor=BrickRed,urlcolor=blue]{hyperref}

\newcommand{\kp}{\mathbf{k}_\parallel}

\def\be{\begin{equation}}      
\def\ee{\end{equation}}
\def\bea{\begin{eqnarray}}      
\def\eea{\end{eqnarray}}

\def\v#1{\mathbf{#1}}
\def\kp {\mathbf{k}_\parallel}

\def\ket#1{\vert #1 \rangle}

\begin{document}

\title{Weyl nodes in periodic structures of superconductors and spin active
  materials}

\author{Ahmet Keles}
\affiliation{Department of Physics and Astronomy, University of Pittsburgh,
  Pittsburgh, PA 15260}
\affiliation{Department of Physics and Astronomy, George
  Mason University, Fairfax, VA 22030}
\author{Erhai Zhao}
\affiliation{Department of Physics and Astronomy, George
  Mason University, Fairfax, VA 22030}

\begin{abstract}
Motivated by recent progress in epitaxial growth of proximity structures
of $s$-wave superconductors (S) and spin-active materials (M), we show that
the periodic structure of S and M can behave effectively as a superconductor
with pairs of point nodes, near which the low energy excitations are Weyl
fermions.
A simple toy model, where M is described by a Kronig-Penney potential with
both spin-orbit coupling and exchange field, is proposed and solved to obtain
the phase diagram of the nodal structure, the spin texture of the Weyl
fermions, as well as the zero energy surface states in the form of open Fermi
lines (``Fermi arcs").
Going beyond the simple model, a lattice model with alternating layers of S
and magnetic $Z_2$ topological insulators (M) is solved. The calculated
spectrum confirms previous prediction of Weyl nodes based on tunneling
Hamiltonian of Dirac electrons.
Our results provide further evidence that periodic structures of S and M are
well suited for engineering gapless topological superconductors.
\end{abstract}

\maketitle

The time-honored recipe for discovering new superconductors with interesting
pairing symmetries or topological properties is via the synthesis of new
materials. In recent years, an alternative approach has been advocated and
gained experimental success. It is based on making proximity structures of
$s$-wave superconductor (referred to as S hereafter) and spin active materials
(M) such as semiconductors, topological insulators \cite{Kane-rmp,Qi2011} or
ferromagnets with spin-orbit and/or exchange coupling. With proper design, the
proximity structure can behave effectively as a superconductor with the
desired symmetry or topology, at energies below the bulk superconducting gap
of S. For example, Fu and Kane \cite{KaneFu} showed that the interface of a
three-dimensional topological band insulator (TI) and an $s$-wave
superconductor is analogous to a spinless $p_x+ip_y$ superconductor that hosts
Majorana zero modes at vortex cores. Similar states also arise in proximity
structures of S and two dimensional electron gas with Rashba spin-orbit
coupling and Zeeman splitting either due to a nearby ferromagnetic insulator
or an external magnetic field
\cite{dasSarma1,dasSarma2,jason,gilRafael,potter}.
In one dimension, e.g., a semiconductor nanowire \cite{Mourik25052012} or a
chain of ferromagnetic atoms deposited on a superconductor
\cite{Nadj-Perge31102014}, Majorana zero modes form at the sample edges. While
it remains a challenge to fabricate and control the interface properties, or
detect the unequivocal experimental signatures of these states, significant
experimental progress has been made in recent years (for an overview, see
Refs. \onlinecite{beenaker,Jason-12}).

In this paper, we explore the possibility of realizing gapless topological
superconductivity in S-M proximity structures. Specifically, we focus on
superconducting states with topologically protected point nodes, i.e., the
analogs of the A phase of superfluid $^3$He \cite{volovik2009universe} and
Weyl semimetals \cite{PhysRevB.83.205101,ying-11,carp-12}. Such a state was
predicted to appear in the superlattice structures of superconductors and
magnetic topological insulators by Meng and Balents, and referred to as ``Weyl
superconductors" \cite{balents1}. The elegant analysis of Ref.
\onlinecite{balents1} is based on an effective Hamiltonian describing the
tunneling of the helical Dirac electrons (the surface states of TI) across the
layers of TI and S, where the presence of the superconducting pairing
potential and Zeeman field provide a mass to the Dirac electrons.  The
proposal of Meng and Balents can be viewed as the generalization of the
earlier work on Weyl semimetal in the multilayer structures of trivial
insulators and magnetic topological insulators \cite{balents2}. One is then
led to the following questions: is it feasible to realize Weyl superconductors
using materials other than topological insulators; is helical Dirac electron
essential?

We answer these questions by considering a simple, idealized model of S-M
superlattice. Here M stands for a general spin active material with both
spin-orbit coupling and exchange splitting (i.e., both the time reversal and
spatial inversion symmetry are broken). We assume, as in
Ref.~\onlinecite{balents1}, that the M layer is sufficiently thin so that the
suppression of superconductivity is not significant and the superconducting
phase coherence is maintained across the M layers. This motivates us to
approximate the M layers as delta function spin-active potentials, similar to
the well known Kronig-Penney model. The band structure of this model is solved
to illustrate the reconstruction of the low energy spectrum due to the
periodic spin-active potential. We identify parameter regimes where the
spectrum has one pair or two pairs of Weyl nodes.  We discuss the low energy
effective Hamiltonians near the Weyl nodes, and the Fermi line (usually
referred to as ``Fermi arc"~\cite{ashvin}) zero energy surface states that
manifest the nontrivial topological properties of the superconducting state.
This model clearly illustrates that neither TI nor helical Dirac electron is
necessary for Weyl nodes to appear in S-M superlattices.

This simple model can be straightforwardly generalized to treat finite
thickness of the M layer. It however neglects many microscopic details of the
specific materials and the S-TI interface. For the purpose of providing design
parameters for Weyl superconductors, it is also desirable to describe the
periodic structure by tight binding models defined on discrete lattices.  We
present such a lattice model for the S-TI superlattice, in which each unit
cell consists of a few layers of S and another few layers of magnetized TI
with a tunable hopping matrix describing the coupling between the two
materials. We outline a procedure to compute the energy spectrum of the
superlattice, and assess the requirements to realize Weyl superconductors with
one pair of nodes and two pairs of nodes respectively. Along the way, we
briefly review the properties of a single S-TI interface \cite{KaneFu}, and
discuss the relation between the Andreev bound states at a single interface
\cite{erhai2} and the spectrum of multilayer systems. 

\section{A simple model for S-M superlattice}

\begin{figure}[h]
  \centering
  \includegraphics[scale=0.8]{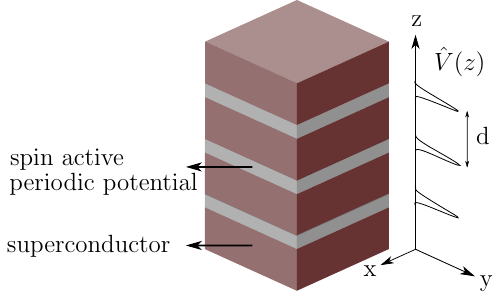}
   \caption{Schematic of the S-M periodic structure.}
   \label{fig:geometry}
\end{figure}

We consider a periodic layered structure of an $s$-wave superconductor (S)
and a spin active material (M) extending in the $z$ direction as schematically
shown in Fig.~\ref{fig:geometry}. As far as the low energy excitations are
concerned (relative to the bulk superconducting gap $\Delta$), M amounts to
a periodic spin-active potential $\hat{{V}}(z+d)=\hat{{V}}(z)$, where the hat
denotes matrices in spin space.  To preserve the superconductivity throughout
the whole structure, the M layers should not be too thick so we assume the M
layer thickness is much smaller than the period $d$. In this limit, $\hat{V}$
can be modeled by Kronig-Penney potential of the form
\be 
  \hat{{V}}(\kp,z) = d\sum_n \delta(z-nd) 
  \big[V_0 +\sum_i {V}_i(\kp) \hat{\sigma}_i \big].
\ee
Here $\hat\sigma$'s are the Pauli matrices in the spin space, $\kp=(k_x,k_y)$
is the transverse momentum {which is conserved due to translational invariance
  on the $xy$ plane}. Note that the material details of M do not enter in this
description, they are encoded in $V_0$ and $V_i$ which are chosen to reproduce
the scattering matrix of electrons by M.  The superlattice is then described
by the following Hamiltonian in the particle-hole space,
\begin{equation}
  \check{\mathcal{H}}({\kp},z) = 
  \left[
    \begin{array}{cc}
      \hat h_0({\kp},z)   &  \Delta i\hat{\sigma}_2  \\
      -\Delta i\hat{\sigma}_2   &  -\hat h_0^*(-{\kp},z)
    \end{array}
  \right],
\end{equation}
with $\hat h_0({\kp},z)=(\kp^2-\partial_z^2)/2m_e -\mu + \hat{{V}}(\kp,z)$,
and the check denotes a matrix in the particle-hole and spin space. Note that
we have assumed $\Delta$ to be homogeneous, and for $z=nd$, i.e. insides M,
the potential $\hat{V}$ dominates all other terms in the Hamiltonian. For
simplicity, we shall put $V_0=0$. To model M with both spin-orbit coupling and
exchange splitting, we assume that $\hat{V}$ takes the form \cite{grein} 
\be
\sum_i {V}_i(\kp) \hat{\sigma}_i = v_{so}(-k_y\hat{\sigma}_x+k_x\hat{\sigma}_y)
  +v_{z}\hat{\sigma}_z,
\ee
where $v_{so}$ is the strength of Rashba spin-orbit coupling, and $v_{z}$ is
the Zeeman (exchange) field along the $z$ direction.
{Using the periodicity of the Hamiltonian,
$\check{\mathcal{H}}(z)=\check{\mathcal{H}}(z+d)$,
the band structure can be easily obtained by the expansion of the wave
function via Bloch's theorem}
\begin{equation}
  \check{\Psi}_{\mathbf{k}}(x,y,z)=
  e^{ik_xx+ik_yy}\sum_Ge^{i(k_z+G)z}\check{\Phi}_{\mathbf{k},G},
  \label{eq:blochwave}
\end{equation}
where $G$ is reciprocal lattice vector $G=m2\pi/d$ and $m\in \mathbb{Z}$.
The generalized Bogliubov-de Gennes (BdG) equation then becomes  
\begin{equation}
  \sum_{G'}\big[ \check{\mathcal{H}}_0 ({\mathbf{k}},G) \delta_{G,G'}
  +\check{\mathcal{V}}(\kp)\big]\check{\Phi}_{\mathbf{k},G'}=E \check\Phi_{\mathbf{k},G}.
  \label{eq:blochhamiltonian}
\end{equation}
Here we have separated the ``unperturbed" Hamiltonian
\begin{equation}
  \check{\mathcal{H}}_0(\mathbf{k}, G) = 
  \left[
    \begin{array}{cc}
      \xi(\kp,k_z+G)   &  \Delta i\hat{\sigma}_2  \\
      -\Delta i\hat{\sigma}_2   &  -\xi(\kp,k_z+G)
    \end{array}
  \right],
\end{equation}
with $\xi(\kp,k_z+G)=(\kp^2+(k_z+G)^2)/2m_e-\mu$, and the spin-active ``perturbation"
\begin{equation}
  \check{\mathcal{V}}(\kp) = v_{so}[-k_y\sigma_x+\tau_z\otimes k_x\sigma_y]
  + v_{z} \tau_z \otimes \sigma_z 
\end{equation}
with $\tau_z$ being the Pauli matrix in the particle-hole space (we will drop
the hat for $\sigma$ when there is no ambiguity).

The infinite dimensional matrix equation in Eq.~\ref{eq:blochhamiltonian} can
be solved numerically by a truncation, keeping only $|G|$ up to some large
enough value of $N\pi/d$, followed by diagonalization to yield the band
dispersion $E_\mathbf{k}$. This truncation is physically equivalent to
introducing a small width to the M layers. Of course one has to check that the
low energy spectrum does not depend on $N$.
This model can easily be generalized to the case of M layers with finite
thickness. For example, {superlattice unit cell can be modeled in a way that
  the region $z\in [0,d_1]$ is occupied with S (where $\Delta$ is constant,
  $\hat{V}$ is zero) and $z\in [d_1, d]$ is occupied with  M (where $\Delta$
  vanishes but $\hat{V}$ is constant)}.  In this case, both $\Delta$ and
$\hat{V}$ have off-diagonal matrix elements in $G$ space and the resulting BdG
equation is slightly more complicated than Eq.~\ref{eq:blochhamiltonian}. 

We are particularly interested in the zero energy solutions of
Eq.~\ref{eq:blochhamiltonian}. For
this purpose, it is useful to introduce $\check\phi_\mathbf{k}
=\sum_G\check{\Phi}_{\mathbf{k},G}$ which can be shown to satisfy the
following equation; 
\begin{equation}
  \check{A}_{\mathbf{k},E}\check\phi_\mathbf{k}\equiv \big[
  1-\sum_G ({E -\check{\mathcal{H}}_0(\mathbf{k},G)})^{-1}\check{\mathcal{ V }}(\kp)
  \big]\check\phi_\mathbf{k} = 0.
  \label{eq:nodes_condition1}
\end{equation}
The matrix inverse is the bare Green function  of the bulk superconductor and
can be computed analytically. Existence of the zero energy solutions ($E=0$)
at isolated nodal points on the $k_z$ axis with $\kp=0$ is equivalent to the
existence of a non-trivial solution of Eq.~\ref{eq:nodes_condition1} which can
be expressed as $\mathrm{det}A_{\mathbf{k},0}=0$. This equation can be further
manipulated analytically to give the following simple equation as the
condition of the zero energy solutions;

\begin{equation}
  g^2+f^2-1=0,
\end{equation}
where $g=v_z\sum_m\xi_m/(\xi_m^2+\Delta^2)$,
$f=v_z\sum_m\Delta/(\xi_m^2+\Delta^2)$ and $\xi_m=(k_z+2\pi m/d)^2/2m_e-\mu$.
This offers a fast way to scan for Weyl nodes in the parameter space since no
numerical diagonalization is required.

Before we present any numerical results, it is worthwhile to develop a
qualitative picture of the low energy excitations in such S-M superlattices.
Consider the normal state dispersion (turning off superconductivity by setting
$\Delta$ to zero) in the absence of M. For $\kp=0$, the low energy excitations
are located at large momenta around $k_z\sim \pm k_F$.  In the presence of a
weak periodic potential $\hat{V}$, the spectrum of the superlattice structure
can be obtained by folding the free dispersion into the first Brillouin zone
$k_z\in[-\pi/d,\pi/d]$ which gives a set of Bloch bands, $(k_z+G)^2/2m_e$, all
being restricted to small momenta since $\pi/d\ll k_F$.  The spectrum acquires
a gap as $\Delta$ is turned on. A finite Zeeman field will split the Andreev
bound states formed below the superconducting gap $\Delta$, and push one of
the branch towards the zero energy.  A non-zero spin-orbit coupling can in
principle endow a topologically nontrivial spin structure to these zero energy
states.
The S-M proximity structure considered here differs from the well studied
system of semiconductor nanowires \cite{jason} in its dimensionality. In three dimensions,
linear dispersion in the vicinity of the nodes is known to be described by
the Weyl Hamiltonian \cite{volovik2009universe,ashvin}.

\section{Weyl nodes}
    
\begin{figure}[h]
  \includegraphics[width=.5\textwidth]{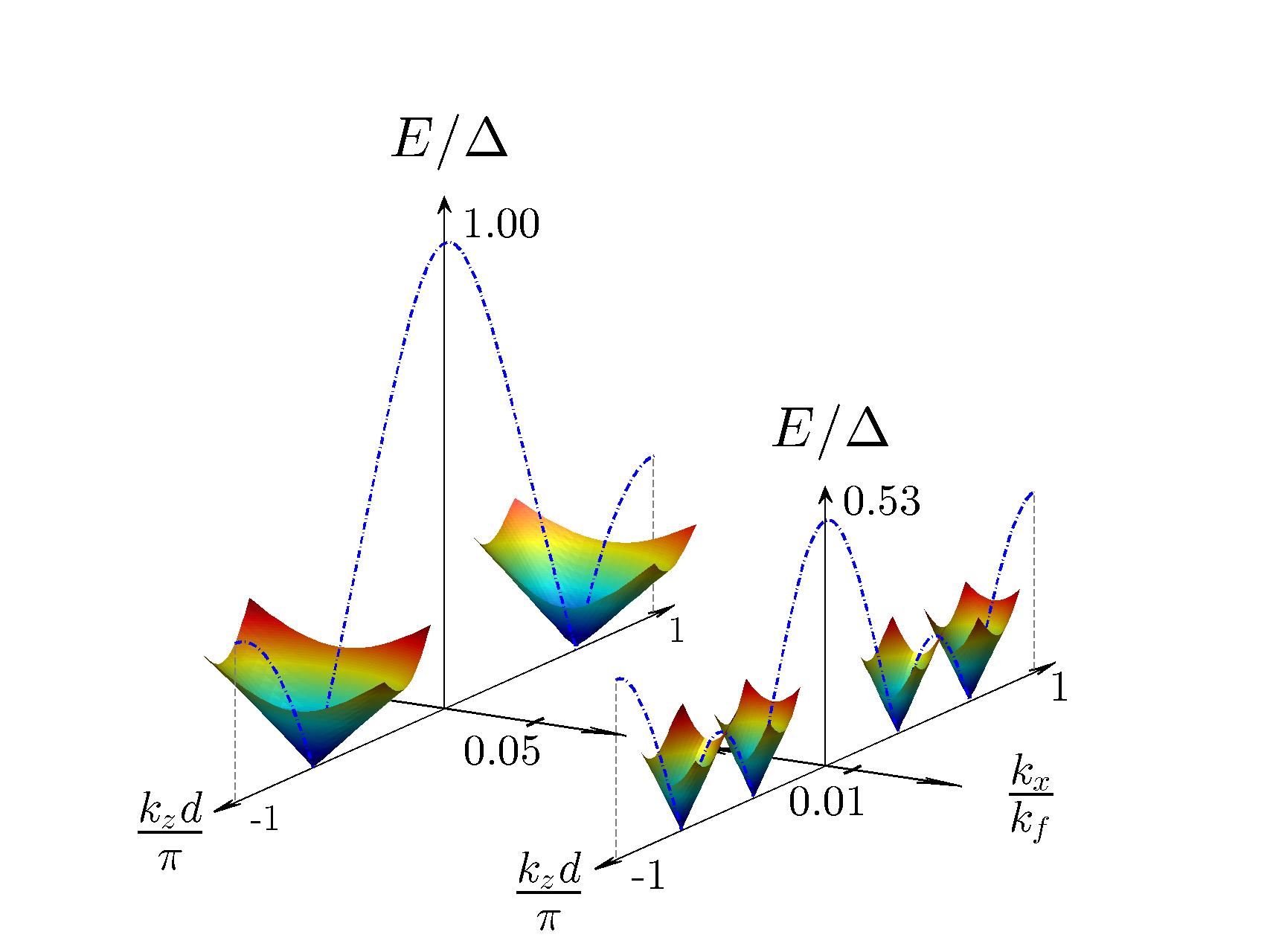}
  \caption{Weyl nodes in S-M superlattices. Only the lowest eigen energy
    $E(k_x,k_y=0,k_z)$ is shown.  Blue line is its dispersion along the $k_z$
    axis inside the first Brillouin zone $k_z\in[-\pi/d,\pi/d]$.  Left panel:
    single pair of Weyl nodes. The range of $k_x$ is $k_x/k_F\in [-.05,.05]$.
    $d = 10\pi/k_F$, $v_z = 2.5\Delta$.  Right panel: two pairs of Weyl node
    for $k_x/k_F\in [-.015,.015]$, $d = 9.5\pi/k_F$, $v_z = \Delta$.  Here $d$
    is the superlattice period, $v_z$ is the strength of Zeeman field and
    $v_{so}$ is the spin-orbit coupling. We choose $\Delta = 0.05\mu$ and
    $v_{so} = 5\Delta$.
  }
  \label{fig:nodes}
\end{figure}

Two representative examples of the low energy spectrum of the S-M superlattice
are shown in Fig.~\ref{fig:nodes}.  We will explicitly show that they
correspond to one pair and two pairs of Weyl nodes, respectively.  In both
cases, we observe a linear energy-momentum dispersion near isolated points
$\{\mathbf{k}^0\}$ located on the $k_z$ axis. At any of these diabolical
points, the zero energy state is doubly degenerate. Let us label these two
degenerate states as $|\check{\Psi}_+\rangle$ and $|\check{\Psi}_-\rangle$, or
$|\pm\rangle$ for short. The low energy physics near the node is described by
an effective Hamiltonian which is a $2\times 2$ matrix in the Hilbert space
spanned by $|\pm\rangle$ with the general form \cite{ashvin}
\be
\mathscr{H} (\mathbf{q})= 
\frac{1}{2}\mathbf{h}(\mathbf{q})\cdot \tilde{\boldsymbol{\tau}}=
\sum_{i,j=1}^{3} {v}_{ij} q_j \tilde{\tau}_i ,\;\;\; 
\mathbf{q}=\mathbf{k}-\mathbf{k}^0.
\ee
Here $\tilde{\tau}_i$ are the pseudospin Pauli matrix in the $\pm$ space. The
low energy Bogoliubov quasiparticles thus resemble massless chiral fermions
(Weyl fermions). The chirality refers to the locking of the pseudospin with
respect to the momentum direction as described by $\mathbf{h}(\mathbf{q})$.
The direction of $\mathbf{h}$, $\hat{h}(\mathbf{q})$, constitutes a mapping
from a sphere in the momentum space enclosing the Weyl nodes to an unit sphere
in pseudospin space. The topological invariant for this mapping is the winding
number \cite{volovik2009universe}
\be
\mathscr{N}=\frac{1}{8\pi}\sum_{i,j,k}\epsilon_{ijk}\int d\Omega_k 
\hat{h}\cdot(\partial_i \hat{h}\times \partial_j\hat{h})
\ee
where the integration is over a closed volume containing $q=0$, and
$\partial_j=\partial/\partial q_j$. In the simplest example $\mathbf{h}=\pm
\mathbf{q}$, $\mathscr{N}=\pm 1$; and for $v_{ij}=\lambda_i\delta_{ij}$,
$\mathscr{N}=\Pi_i \mathrm{sign}(\lambda_i)$. Since $\mathscr{H}$ resembles to
a spin-$1/2$ particle in magnetic field, one can define Berry connection and
the corresponding flux density \cite{ashvin}. Then a Weyl node corresponds to
a magnetic monopole with charge $\mathscr{N}$ in the momentum space. For
periodic systems, the net flux through the  Brillouin zone (BZ) and the net
magnetic charge inside the BZ must be zero \cite{ashvin}. Therefore Weyl nodes
always appear in pairs of opposite charge $\mathscr{N}$. They are well
separated in $\mathbf{k}$ space and topologically stable \cite{ashvin}.
 
\begin{figure}[h]
  \includegraphics[width=.48\textwidth]{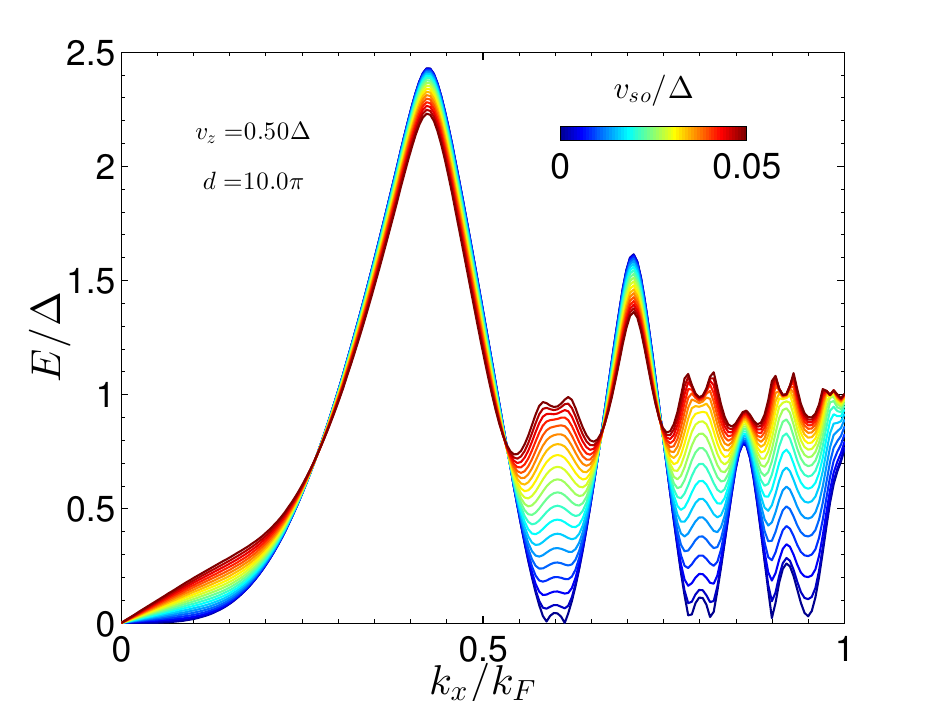}
  \caption{The dispersion of energy with respect to $k_x$ at the Weyl nodes
    for different values of spin-orbit coupling $v_{so}$. The other parameters
    are identical to the left panel of Fig.~\ref{fig:nodes}. 
  }
  \label{kx-disp}
\end{figure}

It is important to conduct the search for zero energy states in the entire
momentum space.  Fig.~\ref{kx-disp} shows the energy-momentum dispersion along
the $k_x$ axis for fixed $k_z=k^0_z$ for various values of $v_{so}$. For very
small spin-orbit coupling, the slope around $k_x=0$ is vanishingly small, and
there are many other low lying states at larger values of $k_x$ that are
sufficiently close to zero energy. These are Andreev bound states (ABS)
ubiquitously found in superconducting proximity structures. For example, for
$v_{so}=0$, the structure reduces to a superconductor-ferromagnet (S-F)
superlattice. It is then expected from semiclassical consideration that a
series of ABS with finite $k_x$ will be formed between two adjacent F layers.
The slope near $k_x=0$ increases with $v_{so}$. At the same time, other ABS at
finite $k_x$ are increasingly gapped out. For large enough $v_{so}$, the Weyl
nodes on the $k_z$ axis are the only zero energy states.
In other words, spin-orbit coupling is crucial for the S-M superlattice to
qualify as a Weyl superconductor. Strong spin-orbit coupling is preferred
because it gives a steep dispersion around the node, making it well separated
from other sub-gap excitations.
    
\begin{figure}[h]
  \includegraphics[width=.23\textwidth]{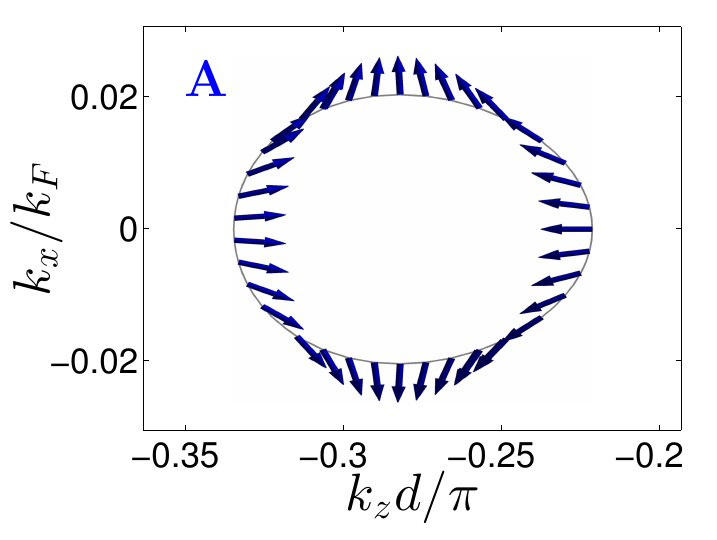}
  \includegraphics[width=.23\textwidth]{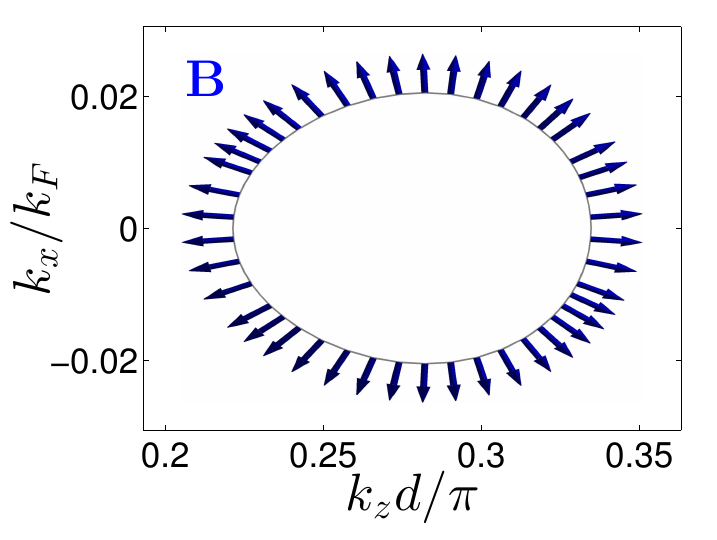}
  \caption{  The pseudospin $\mathbf{h}$ (arrows) on equal energy contours
    (grey lines) near a pair of Weyl nodes. Note that $k_y=0$ and $h_x=0$. To
    show the (three dimensional) $\mathbf{h}$ vector on the $(k_x,k_z)$ plane,
    we plot $h_z$ along the $k_z$ axis and $h_y$ along the $k_x$ axis. Panel A
    is near the node at $-k^0_z$ whereas panel B is near the node at $k^0_z$.
    $\Delta = 0.05\mu$, $v_{so}=0.5\mu$, $v_z = 0.05\mu$, $d/\pi = 10/k_F$.}
  \label{spin2}
\end{figure}

\begin{figure}[h]
  \includegraphics[width=.23\textwidth]{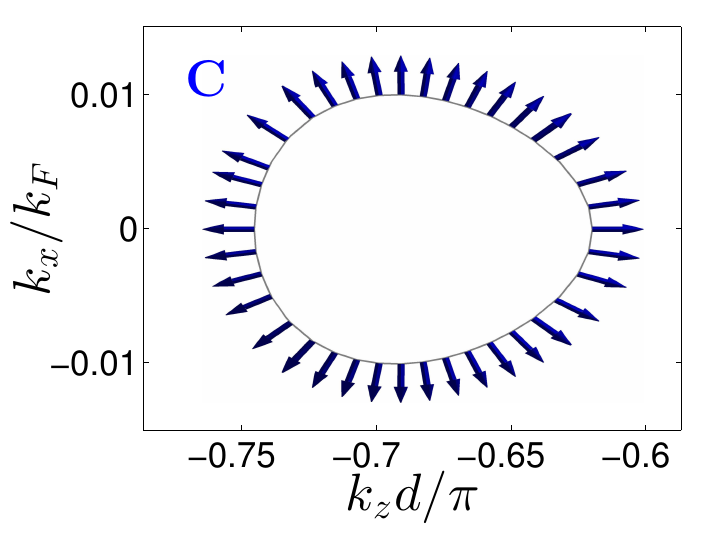}
  \includegraphics[width=.23\textwidth]{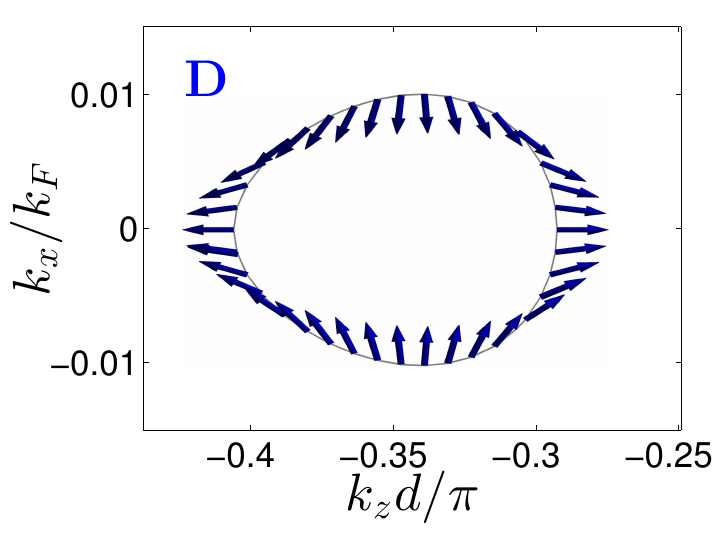}
  \includegraphics[width=.23\textwidth]{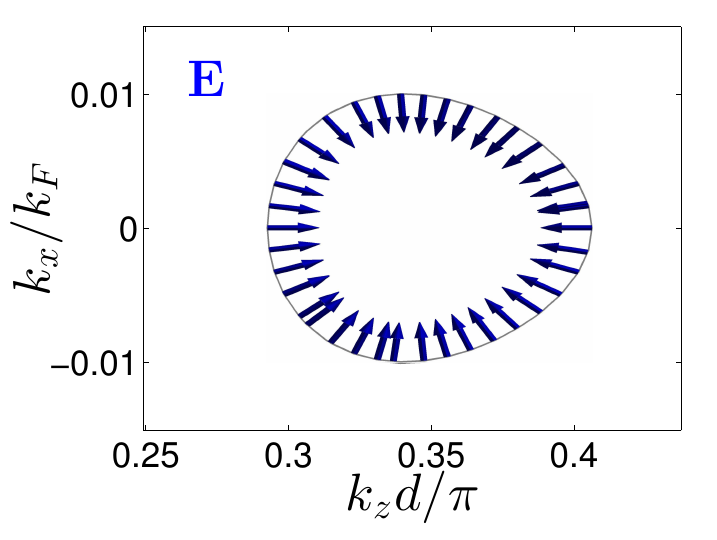}
  \includegraphics[width=.23\textwidth]{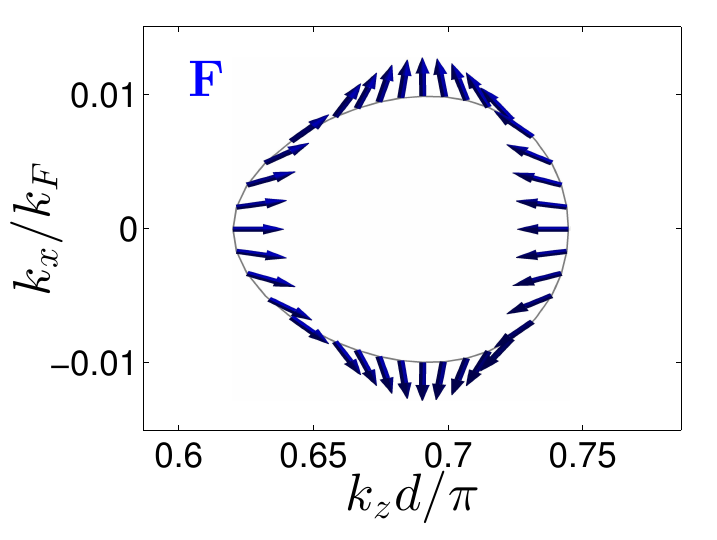}
  \caption{ The pseudospin $\mathbf{h}$ on equal energy contours near two
    pairs of Weyl nodes, C, D, E, and F. $k_y=0$ and $h_x=0$, $h_z$ is along
    the $k_z$ axis and $h_y$ is along the $k_x$ axis. $\Delta = 0.05\mu$,
    $v_{so}=0.5\mu$, $v_z = 0.05\mu$, $d/\pi = 9.5/k_F$. }
  \label{spin4}
\end{figure}

We have numerically computed the pseudospin texture near the Weyl nodes.
Consider a state $|\check{\Psi}_\mathbf{k}\rangle$ with positive energy and a
momentum $\mathbf{k}$ on one of the cones shown in Fig.~\ref{fig:nodes}, and
construct a spinor $\chi$ by projecting $|\check{\Psi}_\mathbf{k}\rangle$ onto the $|\check\Psi_\pm\rangle$ basis formed
by the zero energy states at the given node,
\begin{equation}
  |\chi\rangle=
  \left(
    \begin{array}{c}
      u \\
      v    \\
    \end{array}
  \right)
  \equiv
  \left(
    \begin{array}{c}
      \langle\check{\Psi}_+|\check{\Psi}_\mathbf{k}\rangle    \\
      \langle\check{\Psi}_-|\check{\Psi}_\mathbf{k}\rangle    \\
    \end{array}
  \right).
\end{equation}
The three components of the $\mathbf{h}$ vector are the expectation values of
the corresponding Pauli matrices $\tilde{\tau}_i$ in state $|\chi\rangle$,
\begin{align}
  \mathrm{h}_z &= \frac{1}{2}\left( |u|^2-|v^2|\right),\\
  \mathrm{h}_x &= \mathrm{Re}\left( u^*v\right),\\
  \mathrm{h}_y &= \mathrm{Im}\left( u^*v\right).
\end{align}
The result is illustrated in Fig. \ref{spin2} for the case of single pair of
Weyl nodes.  The pseudospin texture suggests that
\be
\mathscr{H}_\pm=  v_\parallel [q_x \tilde{\tau}_y +q_y \tilde{\tau}_x] \pm v_z q_z \tilde{\tau}_z
\ee
for the two nodes at $\pm k^0_z$ respectively. Using the formula for
$\mathscr{N}$ above, we can rescale $q$ to find their corresponding
topological charge $\mathscr{N}=\pm 1$, i.e., the two Weyl nodes have opposite
topological charge (chirality).  The characteristic spin texture around two
pairs of Weyl nodes is illustrated in Fig.  \ref{spin4}. We observe that the
topological charges of the two nodes on the positive $k_z$ axis, such as $E$
and $F$, are opposite of each other.  The two nodes at mirroring position $\pm
k_z^0$, e.g., $D$ and $E$, also possess opposite charge.

\section{Fermi Arc}

The Andreev bound states formed at the surface of a Weyl superconductor are
very peculiar.  The zero energy surface states in the two dimensional momentum
space (such as the surface Brillouin zone) take the shape of a continuum line
connecting two Weyl nodes of opposite topological charge. Its topological
origin has been discussed in the context of Weyl semimetal
\cite{PhysRevB.83.205101,ashvin} and the A phase of $^3$He
\cite{PhysRevB.83.094510,PhysRevB.86.214511} and will not be repeated here.
Instead we explicitly compute the spectrum of S-M superlattice structures that
have a finite width $L$ in the $y$-direction to demonstrate the Fermi lines.
Since the wave function has to vanish at $y=0$ and $y=L$, we can expand it in
sine Fourier series,
\begin{equation}
  \check{\Psi}(x,y,z) = e^{ik_xx+ik_zz}\sum_{G,n}
  \check{\Phi}_{G,n}e^{iGz}\sin(\frac{n\pi}{L}y).
\end{equation}
Then the BdG equation becomes
\begin{equation}
  \check{\mathcal{H}}_0(G,n)\check\Phi_{G,n}
  +\sum_{n',G'}\check{\mathcal{V}}_{nn'}\check{\Phi}_{G'n'}
  =\epsilon\check{\Phi}_{G,n}
  \label{eq:blochhamiltonianArc}
\end{equation}
where we have defined 
\begin{equation}
  \check{\mathcal{H}}_0(G,n) = 
  \left[
    \begin{array}{cc}
      \xi(k_x,n,k_z+G)   &  \Delta i\sigma_2  \\
      -\Delta i\sigma_2   &  -\xi(k_x,n,k_z+G)
    \end{array}
  \right]
\end{equation}
with $\xi(k_x,n,k_z+G)= (k_x^2 + k_n^2 + (k_z+G)^2)/2m_e -\mu$, $k_n =
n\pi/L$, and $\check{\mathcal{V}}_{nn'}(\mathbf{p}) = \delta_{nn'}
[-v_{so}k_x\sigma_y + v_z\sigma_z]\otimes \tau_3 +v_{so}\kappa_{nn'} \sigma_x$
with 
\begin{align}
  \kappa_{nn'}= \frac{2}{L}\int_0^L dy 
  \sin\left( k_n y\right)
  \left[
    -i\partial_y
  \right]
  \sin\left( k_n' y\right).
\end{align}
In the numerical solution of the BdG equation,
Eq.~\ref{eq:blochhamiltonianArc}, it is important to keep enough $k_n$.

\begin{figure}[h]
  \includegraphics[width=.48\textwidth]{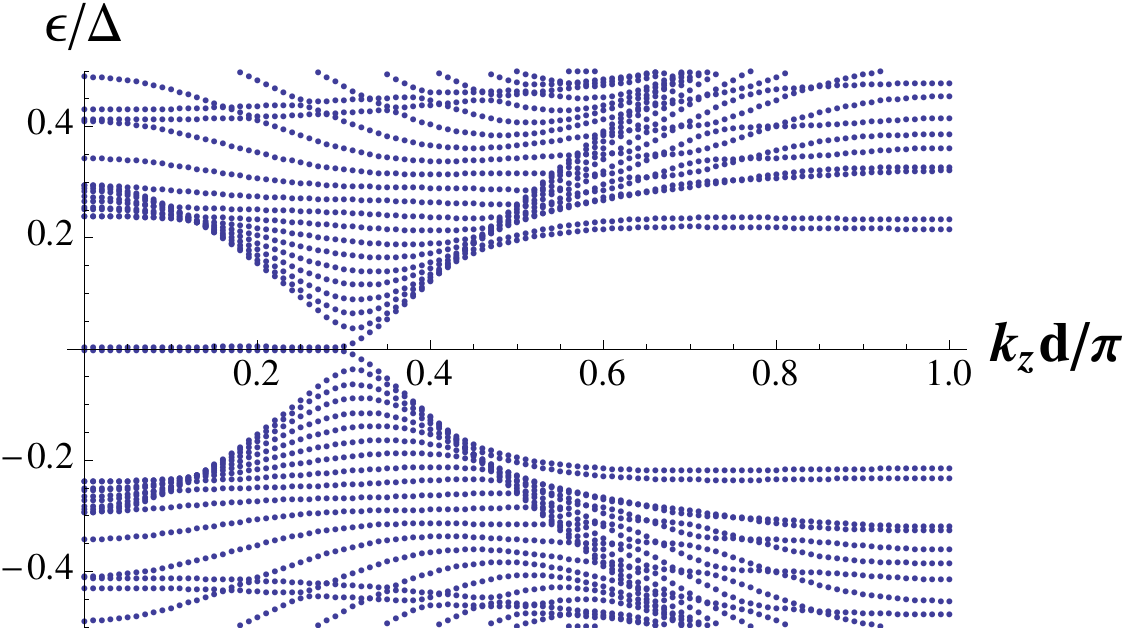}
  \includegraphics[width=.48\textwidth]{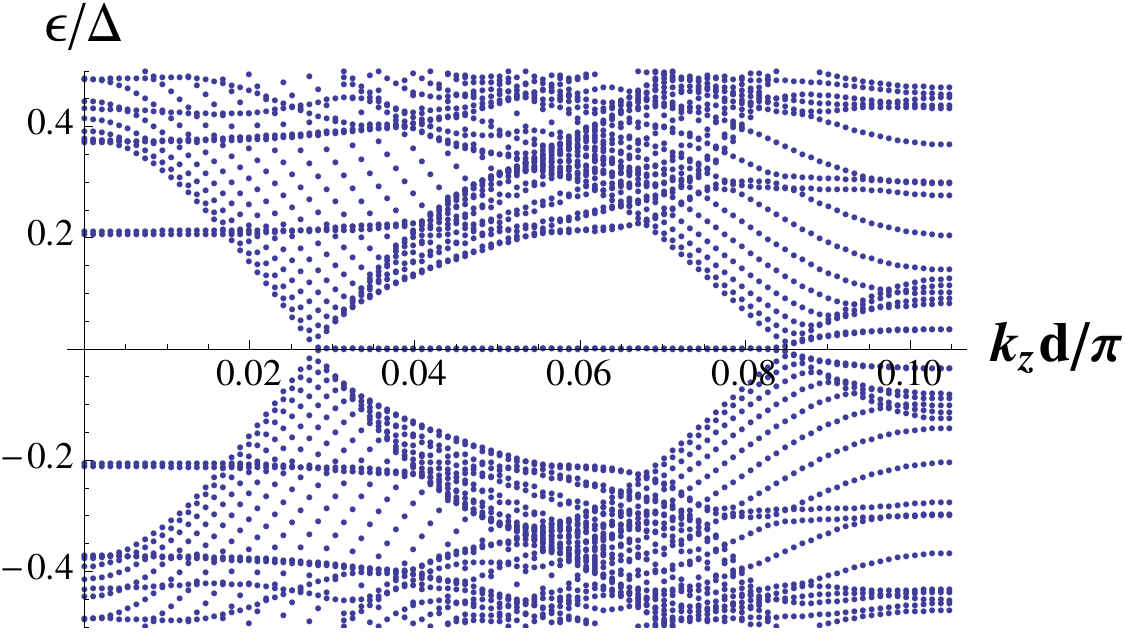}  
  \caption{Fermi arcs on the surface of the S-M superlattice. The plot shows
    the low energy spectrum $\epsilon(k_x=0,k_z)$ of a finite slab
    $y\in[0,L]$.  Top panel: the arc connects $k^0$ and $-k^0$ (only the
    positive $k_z$ is shown).  $k_Fd=50$, $v_z=0.6\Delta$.  Bottom panel: the
    Fermi arc along the $k_z$ axis connects two Weyl nodes on the positive
    $k_z$ axis.  $k_Fd=30$, $v_z=1.1\Delta$. In both cases,  $k_FL=200$,
    $\mu=20\Delta$, $v_{so}=\Delta$. }
  \label{arc}
\end{figure}

Fig. \ref{arc} shows the calculated finite-slab spectrum for single pair
(upper panel) and two pairs (lower panel) of Weyl nodes.  In the former case,
a continuous line of zero energy states is formed along the $k_z$ axis
connecting the bulk Weyl node at $k^0_z$ to that at $-k^0_z$. 
These Fermi arc (in this case just a straight line) states are absent in the
bulk spectrum and correspond to the surface states at $y=0$ and $L$. In fact,
a very small gap is visible due to the hybridization of the two surfaces that
are of finite distance $L$ apart.   
In the latter case, the Fermi arc connects the two bulk Weyl nodes on the
positive $k_z$ axis. (There is another arc along the negative $k_z$ axis,
which is not shown.) This is in accordance with the topological charge of the
nodes found above from the spin texture.

\section{Phase diagram}

\begin{figure}[h]
  \includegraphics[width=.45\textwidth]{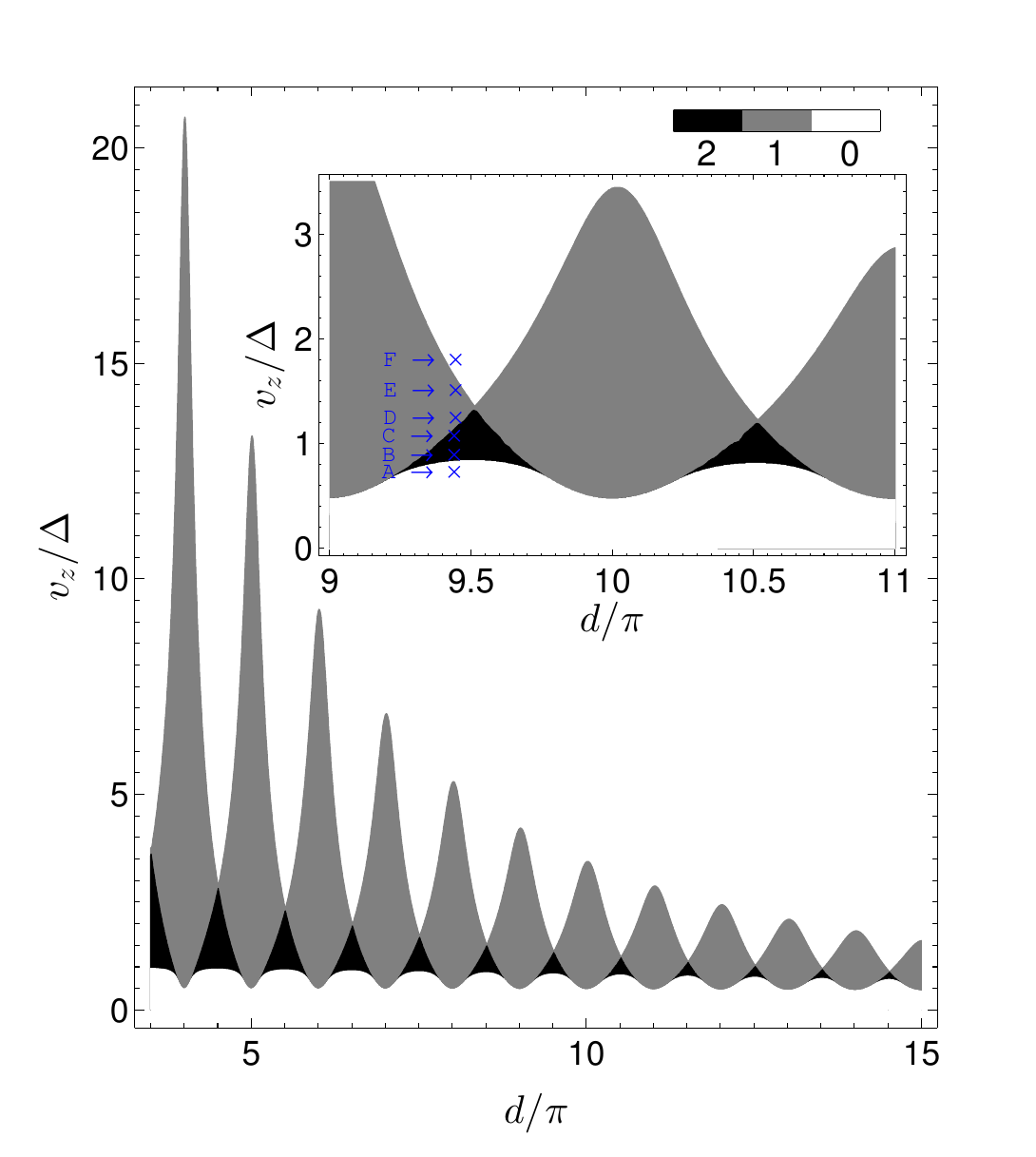}
  \includegraphics[width=.45\textwidth]{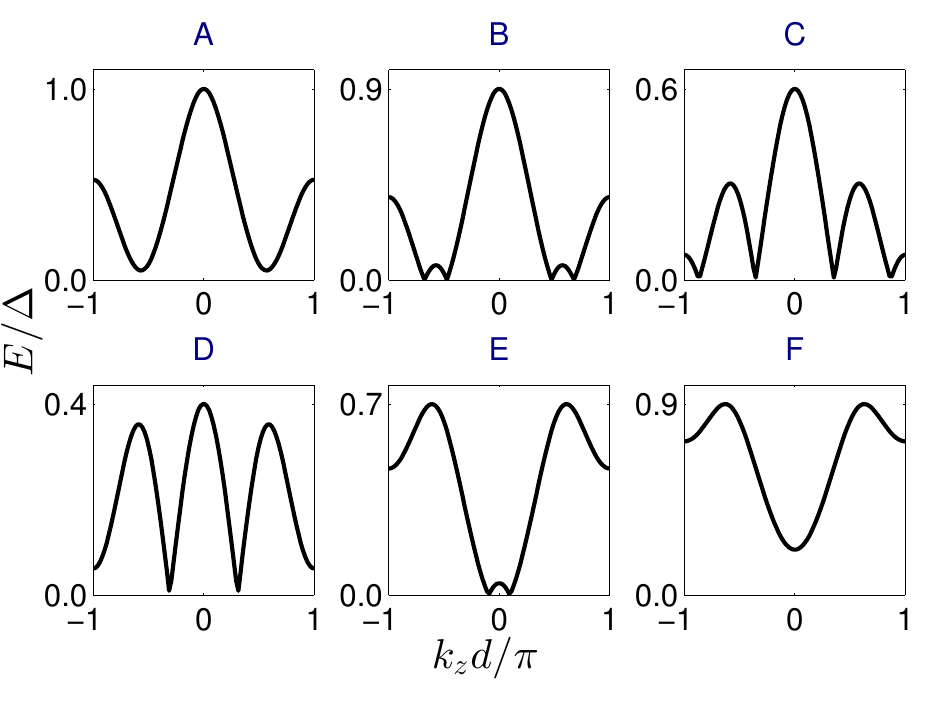}
  \caption{Top: Phase diagram of the nodal structure of the S-M superlattice
    as a function of superlattice spacing $d$ and Zeeman field $v_{so}$ in the
    regime of sufficiently strong spin-orbit coupling.  The shaded (light
    grey) region corresponds to one pair of Weyl nodes, and the dark black
    region corresponds to two pairs of Weyl nodes. The spectrum is gapped in the white
    region.  $\Delta=0.05\mu$, the superlattice period $d$ is measured in
    $1/k_F$.  Inset: zoom-in of the region around $dk_F/\pi=10$.  The energy
    spectra (the lowest energy) along a vertical cut, from point A to point F,
    are shown in the bottom panel.
  }
  \label{phase-diag}
\end{figure}

The spin texture and the Fermi arc surface states presented above
unambiguously established the existence of pairs of Weyl nodes in S-M periodic
structures. We have systematically scanned the parameter space of this model
and the resultant phase diagram is shown in Fig. \ref{phase-diag}. Most
strikingly, one observes a series of lobes (light gray regions,
$\mathbb{W}_1$) that feature single pair of Weyl nodes, similar to that shown
in the left panel of Fig. \ref{fig:nodes}. Roughly speaking, each lobe appears
when $k_Fd=n \pi$ ($n\in \mathbb{Z}$) and $v_z>\Delta$.  The dark black
regions ($\mathbb{W}_2$) represent phases with two pairs of Weyl nodes. They
also appear as a regular array but are well separated from each other and much
smaller in area compared to $\mathbb{W}_1$. In fact, $\mathbb{W}_2$ can be
viewed as the overlapping regions of two adjacent $\mathbb{W}_1$ phases, as
seen in the inset of Fig. \ref{phase-diag} which shows the details near
$d/\pi=10$.  In the rest of the phase diagram (white regions, $\mathbb{W}_0$),
the spectrum is gapped, even through the gap may be numerically small (see for
example the spectrum at point $A$ shown in the first sub-panel).
  
The evolution of the spectrum along a vertical cut in the phase diagram, from
point A to point F, is illustrated in the sub-panels of Fig. \ref{phase-diag}.
As $v_z$ is increased at this particular value of $d$, the lowest branch of
the spectrum is pushed down by the increasing Zeeman splitting to touch $E=0$,
entering the $\mathbb{W}_2$ phase (A$\rightarrow$B). With further increase of
$v_z$, the newly born twin of Weyl nodes with opposite charge become
increasingly detached from each other, with one heading towards $k_z=0$ and
the other towards the BZ boundary. The latter gets gapped out once it reaches
the BZ boundary where it annihilates with its mirror image living on the
negative $k_z$ axis, thus marking the transition from the $\mathbb{W}_2$ phase to the
$\mathbb{W}_1$ phase (C$\rightarrow$D).  Further increase of $v_z$ will push
the spectrum away from $E=0$, and the two Weyl nodes of opposite charge
annihilate with each other at $k_z=0$, leaving behind a vacuum
(E$\rightarrow$F).

\section{Lattice Model}
Now we turn to S-M heterostructures described by tight binding
Hamiltonians defined on discrete lattices. Compared to the continuum model
above, the lattice model can sometimes offer more realistic descriptions of
the material-specific properties, especially regarding the coupling between S
and M, and in the limit where the thickness of S and M is comparable to each other. As a
result, the lattice model can potentially provide more quantitative predictions of the design parameters
of Weyl superconductors. We will illustrate this approach by focusing on the
superconductor-magnetic topological insulator (S-TI) superlattice structure
proposed in Ref.~\onlinecite{balents1}. In contrast to Ref.~\onlinecite{balents1}, 
however, we start from microscopic models of S and TI instead of the low energy 
surface degrees of freedom (i.e., the Dirac electrons).

For simplicity, we model both S and TI on cubic lattice with lattice constant
$a$. Each unit cell of the superlattice consists of $N_S$ layers of S and
$N_T$ layers of TI stacked along the $z$ direction. Let $i$ be the layer
index, the eigenvalue problem has a tri-diagonal structure
\be
T_{i-1,i}\Psi_i + T_{i,i+1}\Psi_{i+1} = (E-H_i)\Psi_i.
\ee
Here $T_{i,i+1}$ is the hopping matrix coupling layer $i$ to the neighboring
layer $i+1$, $H_i$ is the Hamiltonian for the $i$-th layer,
and $\Psi_i$ is the wave function at the $i$-th layer. Note that the
transverse momentum $\kp=(k_x,k_y)$ is conserved. For each S layer,  i.e., 
$i\in [1,N_S]$, 
\begin{equation}
  H_i(\kp)=H_{S}(\kp)=\left(
    \begin{array}{cc}
      \xi(\kp)  &  i\sigma_y  \Delta \\
      -i\sigma_y \Delta  &   -\xi^*(-\kp)
    \end{array}\label{fkmodel}
  \right),
\end{equation}
where $\xi(\kp)=-2t_s(\cos k_x+\cos k_y)-\mu_s$ and $k$ is measured in units of $1/a$.
The hopping between two adjacent S layers is simply
\be
T_{i,i+1}= \left(\begin{array}{cc}
-t_s  &  0 \\
0  &   t_s
\end{array} 
\right).
\ee
For example, we take $t_s=0.18$eV, and $\mu=-4t_s$.
We consider Bi$_2$Se$_3$ as a prime example of 3D $Z_2$ topological insulators,
and model each TI layer by \cite{erhai1} 
\be
\hat{h}_M(\kp)=m\hat{\Gamma}_0 + a_2\sin k_x  \hat{\Gamma}_1 + a_2\sin k_y\hat{\Gamma}_2+v_z\hat{\sigma}_3\otimes \hat{1},
\ee
where $m(\kp)=M-2b_1+2b_2(\cos k_x+\cos k_y-2)$. We choose the basis
($\ket{p_+\uparrow}$, $\ket{p_+\downarrow}$,
$\ket{p_-\uparrow}$,$\ket{p_-\downarrow}$), where $p_\pm$ labels the
hybridized $p_z$ orbital with even (odd) parity \cite{zhang2009}. The Gamma
matrices are defined as $\hat{\Gamma}_0=\hat{\tau}_3\otimes \hat{1}$,
$\hat{\Gamma}_i=\hat{\tau}_1\otimes \hat{\sigma}_i$, with $\hat{\tau}_i$
($\hat{\sigma}_i$) being the Pauli matrices in the orbital (spin) space.
$v_z$ is the Zeeman splitting for magnetically doped Bi$_2$Se$_3$
\cite{Chen06082010,Chang12042013,Zhang29032013}. The coupling between two
adjacent TI layers is given by
\be
\hat{t}_M = b_1\hat{\Gamma}_0-\frac{i}{2}a_1\hat{\Gamma}_3. 
\ee
The isotropic version of $\hat{h}_M$ and $\hat{t}_M $, with $a_1=a_2$, $b_1=b_2$, was 
proposed by Qi et al as a minimal model for 3D topological insulators \cite{qi-field}.
To mimic Bi$_2$Se$_3$, we set the lattice spacing $a=5.2$\AA, which gives the
correct unit cell volume, and $a_i=A_i/a$, $b_i=B_i/a^2$ for $i=1,2$. The
numerical values of $M$, $A_i$, $B_i$ are given in Ref.
\onlinecite{zhang2009}.
With these parameters, our model yields the correct band gap and surface
dispersion, it also reduces to the continuum $\mathbf{k\cdot p}$ Hamiltonian
(the Bernevig-Hughes-Zhang model) in the small $k$ limit \cite{zhang2009},
aside from a topologically trivial $\epsilon_0(\v{k})$ term.
To describe the superconducting proximity effect, we have to generalize the TI
hamiltonian above into the particle-hole space.  For $i\in [N_S+1,N_S+N_T]$, 
\be
H_i(\kp) = \left(\begin{array}{cc}
\hat{h}_M(\kp)  &  0 \\
0  &   -\hat{h}^{*}_M(-\kp)
\end{array} 
\right),
\ee
and accordingly,
\be
T_{i,i+1}= \left(\begin{array}{cc}
\hat{t}_M  &  0 \\
0  &  -\hat{t}^*_M
\end{array} 
\right).
\ee
Finally, the hopping from S to TI is a $4\times 8$ matrix,
\be
T_{N_S,N_{S+1}}= \left(\begin{array}{cc}
\hat{t}_{SM} (\kp)  &  0 \\
0  &  -\hat{t}_{SM}^*(-\kp)
\end{array} 
\right)
\ee
with 
\be
\hat{t}_{SM} (\kp) = \left(\begin{array}{cccc}
J_+ & 0 & J_- &  0 \\
0  &  J_+ & 0 & J_-
\end{array} 
\right).
\ee
Here $J_{\pm}$ is the overlap integral between the $p$-orbital $p_\pm$ of TI
and the $s$-like orbital of S. For simplicity, we assume the spin is conserved
during the hopping, and from the orbital symmetry, $J_{+}=-J_{-}=J$ where $J$
can be tuned from weak to strong \cite{erhai1}. Small $J$ mimics a large
tunneling barrier between S and TI, while large $J$ describes good contact,
i.e., strong coupling between S and TI. 

A standard lattice Fourier transform from the layer index $i$ to quasi
momentum $k_z$ inside the reduced Brillouin zone, $[-\pi/Na, \pi/Na]$ with
$N=N_S+N_T$, gives each $T_{i,i+1}$ a phase factor $e^{ik_za}$. The
Hamiltonian of one unit cell of the superlattice is a matrix of the size
$4N_S+8N_T$ subject to periodic boundary conditions. Then a numerical
diagonalization yields the band structure $E(k_x,k_y,k_z)$ of the S-TI
superlattice.

\section{Andreev bound states at the S-TI interface}
Before discussing the S-TI multilayer system, it is worthwhile to first
consider the spectral properties of a single S-TI interface which have been
studied extensively since the pioneer work of Fu and Kane \cite{KaneFu}.
Comparison with these known results will serve as a critical check of our
lattice model presented above. It also establishes the connection between the
microscopic model here and the effective model of Fu and Kane for the S-TI
interface \cite{KaneFu}, as well as that of Meng and Balents for the S-TI
superlattice \cite{balents1}. 

The spectrum of a single S-TI interface can be conveniently extracted from our
lattice model by taking various limits. Firstly, by setting $J=0$ and $v_z=0$
(but keeping finite $\Delta$), the low energy spectrum reduces to the surface
states of the topological insulator \cite{}. As the thickness $N_T$ is
reduced, the linear Dirac spectrum acquires a gap due to the hybridization
between two TI surfaces \cite{zhang2009-2}. For example, the gap is around
$0.012$eV when $N_T=8$. 

\begin{figure}[h]
  \centering
  \includegraphics[width=.45\textwidth]{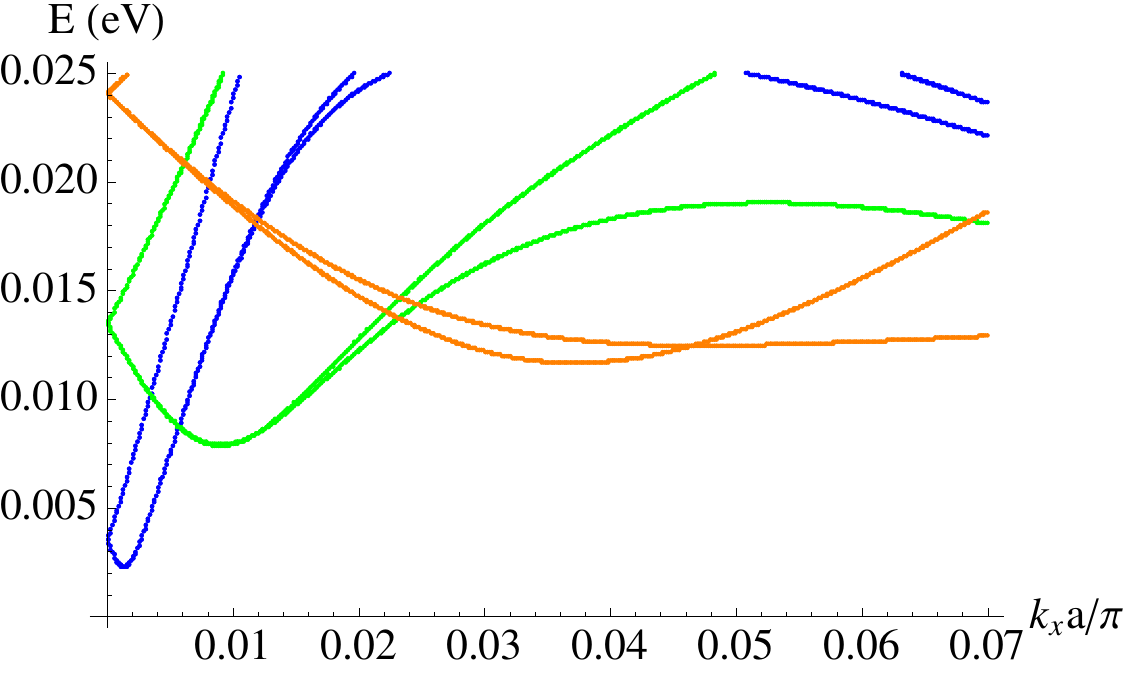}   
  \caption{Andreev bound states at the S-TI interface. $N_S=N_T=20$,
    $k_y=k_z=0$. The blue, green, and red curves are for $J/t_s=0.2$, $0.5$,
    $1$ respectively. Here $v_z=0$, $\Delta=14$meV, $t_s=0.18$eV, and
    $\mu=-4t_s$. }
  \label{decouple}
\end{figure}

Secondly, by setting $v_z=0$ and keeping $N_T$ and $N_S$ large, the low energy
spectrum $E(k_x,k_y,k_z=0)$ reduces to that of a single S-TI interface, since
all the interfaces are sufficiently far apart and essentially decoupled from each other.
Note that in this limit, the BZ is very small, and the dispersion along $k_z$
is negligible so we set $k_z=0$. Fig. \ref{decouple} shows the evolution of
the subgap spectrum as $J$ is increased from the tunneling to the strong
coupling limit. In each case, the dispersion of the sub-gap modes can be fit
well by formula $E(\kp)=\sqrt{\Delta_s^2+(v_s |\kp|\pm \mu_s)^2}$ which
follows from Fu and Kane's phenomenological model \cite{KaneFu}, 
\begin{equation}
H_{FK}(\kp)=\left(
\begin{array}{cc}
h_s(\kp)  &  i\sigma_y  \Delta_s \\
-i\sigma_y \Delta_s  &   -h_s^*(-\kp)
\end{array}
\right),
\end{equation}
where $h_s$ describes the helical Dirac electrons \cite{KaneFu,zhang2009}, 
\be 
h_s(\kp) = -\mu_s + v_s(\sigma_x k_y - \sigma_yk_x).
\ee
This suggests that the Fu-Kane model is valid in a broad range of coupling
strength between S and TI. Yet, as clearly seen in Fig. \ref{decouple}, the
effective parameters $(\mu_s,\Delta_s,v_s)$ in $H_{FK}$ depend sensitively on
$J$. They may get strongly renormalized from their respective nominal values
estimated from the bulk parameters by the proximity effect (the coupling to
S).
For device applications, e.g. for the generation and manipulation of Majorana zero
modes, a large $\Delta_s$ and thus a strong S-TI coupling is preferred. In this
limit, it is more natural to think of the interface state as the Andreev
bound state which penetrates into the superconductor over the coherence length
but decays rapidly (over a distance on the atomic scale) in the TI \cite{erhai2}. 
The lattice calculation presented here is in agreement with the wave function
calculation of Lababidi and Zhao in
Ref.~\onlinecite{erhai2} and the Green function calculation of Grein et al in
Ref. \onlinecite{grein2} .

Thirdly, $v_z$ can be easily incorporated into the Fu-Kane Hamiltonian. For a
single S-TI interface, it opens up a Zeeman gap at $\kp=0$. Using this as the
starting point, Meng and Balents \cite{balents1} analyzed the effective
Hamiltonian of the S-TI superlattice and arrived at a very clean phase diagram
in the plane of $v_z$ and $\Delta$. 

\section{Weyl fermions}
In search of Weyl nodes within our microscopic lattice model, we shall focus
on ``ideal" conditions provided that they seem experimentally feasible. High
temperature superconductor (BSCCO) in proximity to Bi$_2$Se$_3$ was reported
to induce a gap of 15meV, and the pairing symmetry was postulated to be
$s$-wave because no $d$-wave nodes were observed
\cite{Wang2013,Zareapour2012}. In comparison, the gap of Bi$_2$Se$_3$ grown on
NbSe$_3$ is on the order of meV \cite{Wang06042012}. We will consider a fairly
large gap $\Delta=14$meV. The Zeeman field $v_z$ can exceed the value of
$\Delta$. For example, chromium-doped Bi$_2$(Se$_x$Te$_{1-x}$)$_3$ has an
exchange gap of 40meV \cite{Zhang29032013}, and Bi$_2$Se$_3$ doped with Mn
\cite{Xu2012,Chen06082010} develops an exchange gap ranging from 10 to 60 meV.
As to the number of layers for each materials, we will consider for example
$N_S=5$, $N_T=3$ which gives a large BZ, allowing $k_z$ to have significant
dispersion. Certain natural multilayer heterostructures (not superconducting)
of TI such as (PbSe)$_5$(Bi$_2$Se$_3$)$_{3m}$ have been reported
\cite{PhysRevLett.109.236804}. Epitaxial growth of TI films with 1 to 12
quintuple layers on superconducting substrate has been successfully
demonstrated \cite{jia-13}. Also the superconductor BaBiO$_3$ with $T_c\sim
30$K was predicted to turn into a TI upon electron doping \cite{Yan2013}. This
implies that the S-TI superlattice may even be realized based on a single
compound by modulated electric or chemical doping.

\begin{figure}[h]
  \centering
  \includegraphics[width=.4\textwidth]{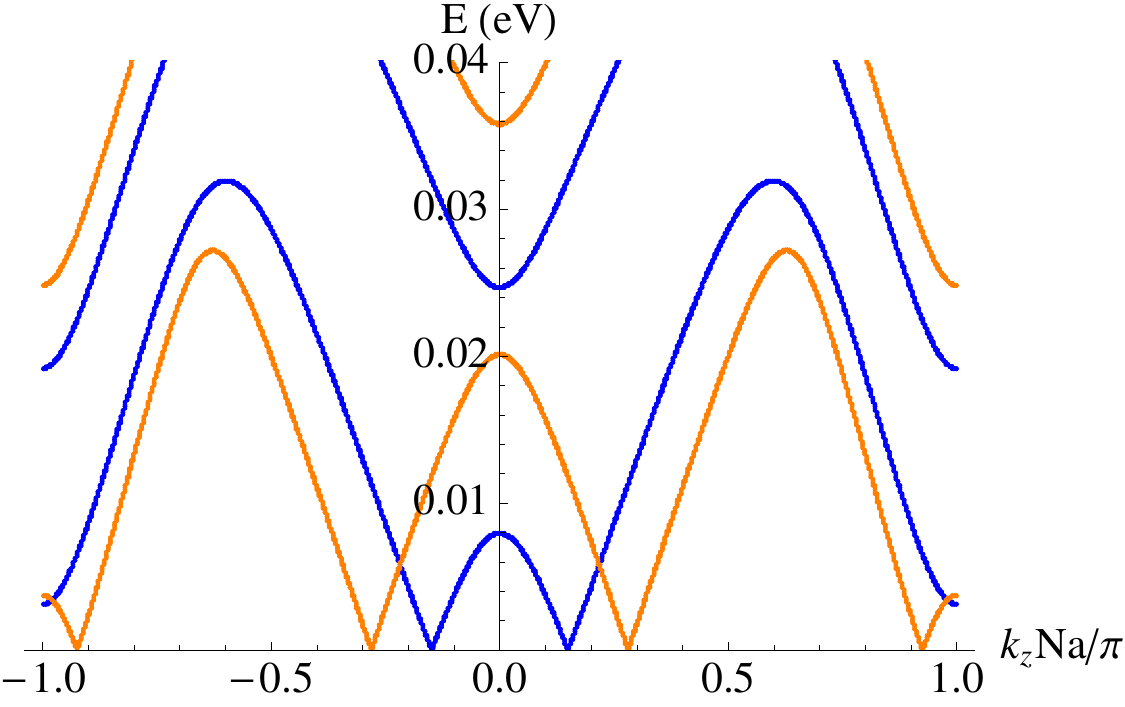}   
  \caption{One pair versus two pairs of Weyl nodes in the S-TI superlattice.
    The blue (orange) spectrum corresponds to $v_{z}=3\Delta$ (5$\Delta$).
    $N_S=5$, $N_T=3$, with $J=0.7 t_s$, $t_s=0.18$eV, $\Delta=14$meV.
  }
 \label{lattice-w}
\end{figure}

\begin{figure}[h]
  \centering
  \includegraphics[width=.4\textwidth]{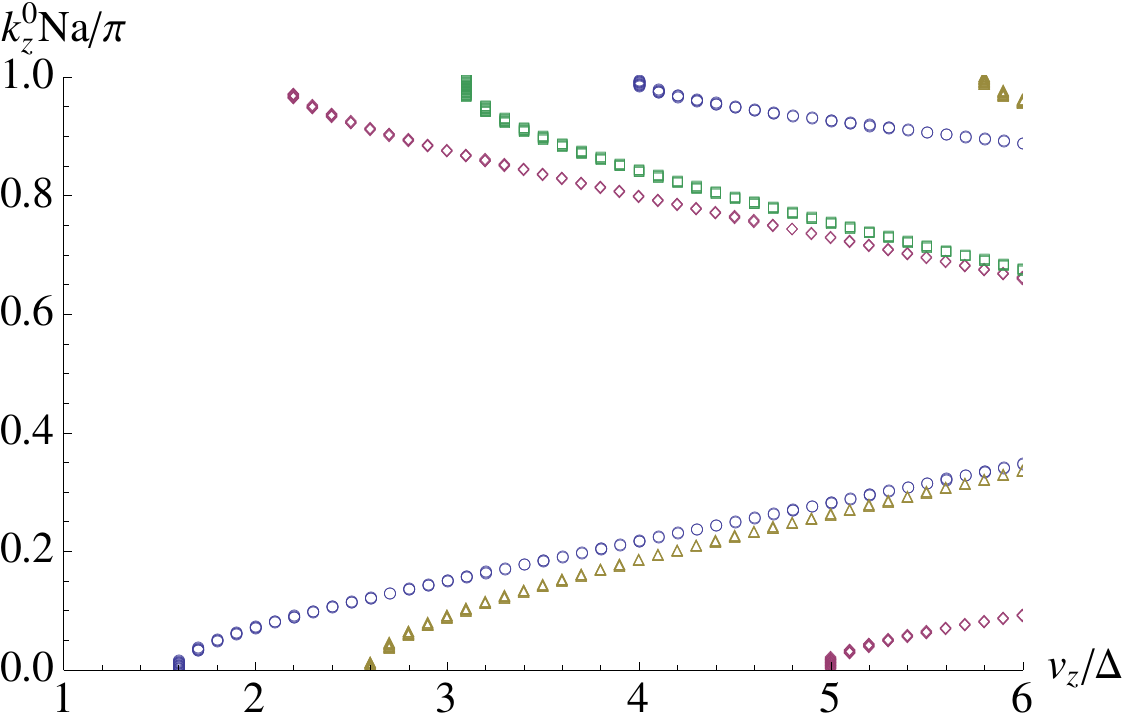}   
  \caption{The number and location of Weyl nodes in the S-TI superlattice as
    functions of the Zeeman field $v_z$.  The circle, diamond, triangle, and
    square are for $N_S=5, 7, 9, 11$ respectively.  $N_T=3$ with $J=0.7 t_s$,
    $t_s=0.18$eV, $\Delta=14$meV.}
   \label{nodal}
\end{figure}

Fig. \ref{lattice-w} shows the low energy part of $E(k_x=k_y=0,k_z)$ for
$v_{z}=3\Delta$ (blue) and 5$\Delta$ (orange). They have one pair and two
pairs of Weyl nodes on the $k_z$ axis respectively. One can explicitly check
that these nodes are the only zero energy states within the reduced BZ, and the
energy $E$ is indeed linear in $\mathbf{k}-\mathbf{k}^0$.
Fig. \ref{nodal} summarizes the location (and number) of the Weyl nodes on the
positive $k_z$ axis as the Zeeman field $v_z$ is increased.  The phases and
phase boundary can be easily read off from Fig. \ref{nodal}. Take $N_S=5$ (the
empty circle) for example, for $v_z<1.6\Delta$, the spectrum is gaped and the
systems is in phase $\mathbb{W}_0$. For $v_z\in[1.6\Delta, 4\Delta]$, there is
only one node on the positive $k_z$ axis. And the node moves away from $k_z=0$ as
$v_z$ is increased. This is phase $\mathbb{W}_1$. For $v_z>4\Delta$, a second
nodal point appears at the BZ boundary $k_z=\pi/(Na)$. The system enters the
$\mathbb{W}_2$ phase. Overall, the evolution of the nodal structure here is similar to that of the
continuum model discussed above.
Fig. \ref{nodal} also compares the phases for increasing number of
superconducting layers. As a general trend, the critical $v_z$ required for
the $\mathbb{W}_{1,2}$ to appear increases with $N_S$. In the limit of large
$N_S$, the dispersion along $k_z$ becomes very flat. 

\section{Concluding remarks}

We have presented two complementary approaches to model and compute the properties of S-M superlattice
structures. The first approach is based on a simple continuum model where M is
described by periodic spin active potentials that are spatially thin compared to the
thickness of the superconductor, and accordingly, the size of the unit cell can be larger than 
the superconducting coherence length. The second approach
is based on a tight binding lattice model describing alternative layers of S
and TI, both of which can be only of several layers thick with a tunable
coupling strength between the two materials. In both models, we find phases that have one pair or two pairs of
Weyl nodes. Together with previous results based on tunneling Hamiltonians
\cite{balents1}, our study unambiguously establishes that (a) S-M periodic
structures can behave as Weyl superconductors at low energies; and (b) for
this to occur, neither topological insulators nor Dirac electrons are
necessary, only the right combination of spin-orbit coupling and Zeeman
splitting are required.
These observations generalize the proposals of realizing gapped topological
superconductors featuring Majorana zero modes in one and two dimensions
\cite{dasSarma1,dasSarma2,jason,gilRafael,potter} to gapless topological
superconductors in three dimensions using periodic structures of S and M. We
hope that the theoretical analysis presented here can stimulate experimental
work to explore these ideas.
 
The emergence of Weyl fermions at low energies out of the vacuum of a conventional $s$-wave superconductor 
is quite striking. While these low energy quasiparticles can be viewed as the Andreev bound states
formed at the S-M interface dispersing with $k_z$ and crossing zero energy at
isolated $\mathbf{k}$ points, they are located in $\mathbf{k}$ space near
$\mathbf{k}=0$ (for $k_Fd\gg 1$), instead of being around the original Fermi
surface $|\mathbf{k}|=k_F$. In other words, the spectral weigh is transferred
not only from high to low energy, but also from high to low momentum, by the
presence of the periodic spin active potential. In this paper we have only considered
one dimensional S-M superlattice structures. It is very likely that  
equally interesting phenomena may rise for ``metamaterials" of
superconductors and spin active materials that have more complicated periodic
structures in even higher dimensions.

\acknowledgements
This work is supported by AFOSR FA9550-12-1-0079 and NSF
PHY-1205504. We thank Matthias Eschrig, Mahmoud Lababidi, and Sungkit Yip for stimulating
discussions. We also acknowledge the Institute for Nuclear Theory at the
University of Washington for its hospitality (INT-15-1 workshop) and the
Department of Energy for partial support during the completion of this work.
\bibliography{RpaV4}

\begin{thebibliography}{38}%
\makeatletter
\providecommand \@ifxundefined [1]{%
 \@ifx{#1\undefined}
}%
\providecommand \@ifnum [1]{%
 \ifnum #1\expandafter \@firstoftwo
 \else \expandafter \@secondoftwo
 \fi
}%
\providecommand \@ifx [1]{%
 \ifx #1\expandafter \@firstoftwo
 \else \expandafter \@secondoftwo
 \fi
}%
\providecommand \natexlab [1]{#1}%
\providecommand \enquote  [1]{``#1''}%
\providecommand \bibnamefont  [1]{#1}%
\providecommand \bibfnamefont [1]{#1}%
\providecommand \citenamefont [1]{#1}%
\providecommand \href@noop [0]{\@secondoftwo}%
\providecommand \href [0]{\begingroup \@sanitize@url \@href}%
\providecommand \@href[1]{\@@startlink{#1}\@@href}%
\providecommand \@@href[1]{\endgroup#1\@@endlink}%
\providecommand \@sanitize@url [0]{\catcode `\\12\catcode `\$12\catcode
  `\&12\catcode `\#12\catcode `\^12\catcode `\_12\catcode `\%12\relax}%
\providecommand \@@startlink[1]{}%
\providecommand \@@endlink[0]{}%
\providecommand \url  [0]{\begingroup\@sanitize@url \@url }%
\providecommand \@url [1]{\endgroup\@href {#1}{\urlprefix }}%
\providecommand \urlprefix  [0]{URL }%
\providecommand \Eprint [0]{\href }%
\providecommand \doibase [0]{http://dx.doi.org/}%
\providecommand \selectlanguage [0]{\@gobble}%
\providecommand \bibinfo  [0]{\@secondoftwo}%
\providecommand \bibfield  [0]{\@secondoftwo}%
\providecommand \translation [1]{[#1]}%
\providecommand \BibitemOpen [0]{}%
\providecommand \bibitemStop [0]{}%
\providecommand \bibitemNoStop [0]{.\EOS\space}%
\providecommand \EOS [0]{\spacefactor3000\relax}%
\providecommand \BibitemShut  [1]{\csname bibitem#1\endcsname}%
\let\auto@bib@innerbib\@empty
%</preamble>
\bibitem [{\citenamefont {Hasan}\ and\ \citenamefont {Kane}(2010)}]{Kane-rmp}%
  \BibitemOpen
  \bibfield  {author} {\bibinfo {author} {\bibfnamefont {M.~Z.}\ \bibnamefont
  {Hasan}}\ and\ \bibinfo {author} {\bibfnamefont {C.~L.}\ \bibnamefont
  {Kane}},\ }\href {\doibase 10.1103/RevModPhys.82.3045} {\bibfield  {journal}
  {\bibinfo  {journal} {Rev. Mod. Phys.}\ }\textbf {\bibinfo {volume} {82}},\
  \bibinfo {pages} {3045} (\bibinfo {year} {2010})}\BibitemShut {NoStop}%
\bibitem [{\citenamefont {Qi}\ and\ \citenamefont {Zhang}(2011)}]{Qi2011}%
  \BibitemOpen
  \bibfield  {author} {\bibinfo {author} {\bibfnamefont {X.-L.}\ \bibnamefont
  {Qi}}\ and\ \bibinfo {author} {\bibfnamefont {S.-C.}\ \bibnamefont {Zhang}},\
  }\href {\doibase 10.1103/RevModPhys.83.1057} {\bibfield  {journal} {\bibinfo
  {journal} {Rev. Mod. Phys.}\ }\textbf {\bibinfo {volume} {83}},\ \bibinfo
  {pages} {1057} (\bibinfo {year} {2011})}\BibitemShut {NoStop}%
\bibitem [{\citenamefont {Fu}\ and\ \citenamefont {Kane}(2008)}]{KaneFu}%
  \BibitemOpen
  \bibfield  {author} {\bibinfo {author} {\bibfnamefont {L.}~\bibnamefont
  {Fu}}\ and\ \bibinfo {author} {\bibfnamefont {C.~L.}\ \bibnamefont {Kane}},\
  }\href {\doibase 10.1103/PhysRevLett.100.096407} {\bibfield  {journal}
  {\bibinfo  {journal} {Phys. Rev. Lett.}\ }\textbf {\bibinfo {volume} {100}},\
  \bibinfo {pages} {096407} (\bibinfo {year} {2008})}\BibitemShut {NoStop}%
\bibitem [{\citenamefont {Sau}\ \emph {et~al.}(2010)\citenamefont {Sau},
  \citenamefont {Lutchyn}, \citenamefont {Tewari},\ and\ \citenamefont
  {Das~Sarma}}]{dasSarma1}%
  \BibitemOpen
  \bibfield  {author} {\bibinfo {author} {\bibfnamefont {J.~D.}\ \bibnamefont
  {Sau}}, \bibinfo {author} {\bibfnamefont {R.~M.}\ \bibnamefont {Lutchyn}},
  \bibinfo {author} {\bibfnamefont {S.}~\bibnamefont {Tewari}}, \ and\ \bibinfo
  {author} {\bibfnamefont {S.}~\bibnamefont {Das~Sarma}},\ }\href {\doibase
  10.1103/PhysRevLett.104.040502} {\bibfield  {journal} {\bibinfo  {journal}
  {Phys. Rev. Lett.}\ }\textbf {\bibinfo {volume} {104}},\ \bibinfo {pages}
  {040502} (\bibinfo {year} {2010})}\BibitemShut {NoStop}%
\bibitem [{\citenamefont {Lutchyn}\ \emph {et~al.}(2010)\citenamefont
  {Lutchyn}, \citenamefont {Sau},\ and\ \citenamefont {Das~Sarma}}]{dasSarma2}%
  \BibitemOpen
  \bibfield  {author} {\bibinfo {author} {\bibfnamefont {R.~M.}\ \bibnamefont
  {Lutchyn}}, \bibinfo {author} {\bibfnamefont {J.~D.}\ \bibnamefont {Sau}}, \
  and\ \bibinfo {author} {\bibfnamefont {S.}~\bibnamefont {Das~Sarma}},\ }\href
  {\doibase 10.1103/PhysRevLett.105.077001} {\bibfield  {journal} {\bibinfo
  {journal} {Phys. Rev. Lett.}\ }\textbf {\bibinfo {volume} {105}},\ \bibinfo
  {pages} {077001} (\bibinfo {year} {2010})}\BibitemShut {NoStop}%
\bibitem [{\citenamefont {Alicea}(2010)}]{jason}%
  \BibitemOpen
  \bibfield  {author} {\bibinfo {author} {\bibfnamefont {J.}~\bibnamefont
  {Alicea}},\ }\href {\doibase 10.1103/PhysRevB.81.125318} {\bibfield
  {journal} {\bibinfo  {journal} {Phys. Rev. B}\ }\textbf {\bibinfo {volume}
  {81}},\ \bibinfo {pages} {125318} (\bibinfo {year} {2010})}\BibitemShut
  {NoStop}%
\bibitem [{\citenamefont {Oreg}\ \emph {et~al.}(2010)\citenamefont {Oreg},
  \citenamefont {Refael},\ and\ \citenamefont {von Oppen}}]{gilRafael}%
  \BibitemOpen
  \bibfield  {author} {\bibinfo {author} {\bibfnamefont {Y.}~\bibnamefont
  {Oreg}}, \bibinfo {author} {\bibfnamefont {G.}~\bibnamefont {Refael}}, \ and\
  \bibinfo {author} {\bibfnamefont {F.}~\bibnamefont {von Oppen}},\ }\href
  {\doibase 10.1103/PhysRevLett.105.177002} {\bibfield  {journal} {\bibinfo
  {journal} {Phys. Rev. Lett.}\ }\textbf {\bibinfo {volume} {105}},\ \bibinfo
  {pages} {177002} (\bibinfo {year} {2010})}\BibitemShut {NoStop}%
\bibitem [{\citenamefont {Potter}\ and\ \citenamefont {Lee}(2011)}]{potter}%
  \BibitemOpen
  \bibfield  {author} {\bibinfo {author} {\bibfnamefont {A.~C.}\ \bibnamefont
  {Potter}}\ and\ \bibinfo {author} {\bibfnamefont {P.~A.}\ \bibnamefont
  {Lee}},\ }\href {\doibase 10.1103/PhysRevB.83.184520} {\bibfield  {journal}
  {\bibinfo  {journal} {Phys. Rev. B}\ }\textbf {\bibinfo {volume} {83}},\
  \bibinfo {pages} {184520} (\bibinfo {year} {2011})}\BibitemShut {NoStop}%
\bibitem [{\citenamefont {Mourik}\ \emph {et~al.}(2012)\citenamefont {Mourik},
  \citenamefont {Zuo}, \citenamefont {Frolov}, \citenamefont {Plissard},
  \citenamefont {Bakkers},\ and\ \citenamefont {Kouwenhoven}}]{Mourik25052012}%
  \BibitemOpen
  \bibfield  {author} {\bibinfo {author} {\bibfnamefont {V.}~\bibnamefont
  {Mourik}}, \bibinfo {author} {\bibfnamefont {K.}~\bibnamefont {Zuo}},
  \bibinfo {author} {\bibfnamefont {S.~M.}\ \bibnamefont {Frolov}}, \bibinfo
  {author} {\bibfnamefont {S.~R.}\ \bibnamefont {Plissard}}, \bibinfo {author}
  {\bibfnamefont {E.~P. A.~M.}\ \bibnamefont {Bakkers}}, \ and\ \bibinfo
  {author} {\bibfnamefont {L.~P.}\ \bibnamefont {Kouwenhoven}},\ }\href
  {\doibase 10.1126/science.1222360} {\bibfield  {journal} {\bibinfo  {journal}
  {Science}\ }\textbf {\bibinfo {volume} {336}},\ \bibinfo {pages} {1003}
  (\bibinfo {year} {2012})}\BibitemShut {NoStop}%
\bibitem [{\citenamefont {Nadj-Perge}\ \emph {et~al.}(2014)\citenamefont
  {Nadj-Perge}, \citenamefont {Drozdov}, \citenamefont {Li}, \citenamefont
  {Chen}, \citenamefont {Jeon}, \citenamefont {Seo}, \citenamefont {MacDonald},
  \citenamefont {Bernevig},\ and\ \citenamefont
  {Yazdani}}]{Nadj-Perge31102014}%
  \BibitemOpen
  \bibfield  {author} {\bibinfo {author} {\bibfnamefont {S.}~\bibnamefont
  {Nadj-Perge}}, \bibinfo {author} {\bibfnamefont {I.~K.}\ \bibnamefont
  {Drozdov}}, \bibinfo {author} {\bibfnamefont {J.}~\bibnamefont {Li}},
  \bibinfo {author} {\bibfnamefont {H.}~\bibnamefont {Chen}}, \bibinfo {author}
  {\bibfnamefont {S.}~\bibnamefont {Jeon}}, \bibinfo {author} {\bibfnamefont
  {J.}~\bibnamefont {Seo}}, \bibinfo {author} {\bibfnamefont {A.~H.}\
  \bibnamefont {MacDonald}}, \bibinfo {author} {\bibfnamefont {B.~A.}\
  \bibnamefont {Bernevig}}, \ and\ \bibinfo {author} {\bibfnamefont
  {A.}~\bibnamefont {Yazdani}},\ }\href {\doibase 10.1126/science.1259327}
  {\bibfield  {journal} {\bibinfo  {journal} {Science}\ }\textbf {\bibinfo
  {volume} {346}},\ \bibinfo {pages} {602} (\bibinfo {year}
  {2014})}\BibitemShut {NoStop}%
\bibitem [{\citenamefont {Beenakker}(2013)}]{beenaker}%
  \BibitemOpen
  \bibfield  {author} {\bibinfo {author} {\bibfnamefont {C.}~\bibnamefont
  {Beenakker}},\ }\href {\doibase 10.1146/annurev-conmatphys-030212-184337}
  {\bibfield  {journal} {\bibinfo  {journal} {Annual Review of Condensed Matter
  Physics}\ }\textbf {\bibinfo {volume} {4}},\ \bibinfo {pages} {113} (\bibinfo
  {year} {2013})}\BibitemShut {NoStop}%
\bibitem [{\citenamefont {Alicea}(2012)}]{Jason-12}%
  \BibitemOpen
  \bibfield  {author} {\bibinfo {author} {\bibfnamefont {J.}~\bibnamefont
  {Alicea}},\ }\href {http://stacks.iop.org/0034-4885/75/i=7/a=076501}
  {\bibfield  {journal} {\bibinfo  {journal} {Reports on Progress in Physics}\
  }\textbf {\bibinfo {volume} {75}},\ \bibinfo {pages} {076501} (\bibinfo
  {year} {2012})}\BibitemShut {NoStop}%
\bibitem [{\citenamefont {Volovik}(2009)}]{volovik2009universe}%
  \BibitemOpen
  \bibfield  {author} {\bibinfo {author} {\bibfnamefont {G.}~\bibnamefont
  {Volovik}},\ }\href@noop {} {\emph {\bibinfo {title} {The universe in a
  helium droplet}}}\ (\bibinfo  {publisher} {Oxford University Press New
  York},\ \bibinfo {year} {2009})\BibitemShut {NoStop}%
\bibitem [{\citenamefont {Wan}\ \emph {et~al.}(2011)\citenamefont {Wan},
  \citenamefont {Turner}, \citenamefont {Vishwanath},\ and\ \citenamefont
  {Savrasov}}]{PhysRevB.83.205101}%
  \BibitemOpen
  \bibfield  {author} {\bibinfo {author} {\bibfnamefont {X.}~\bibnamefont
  {Wan}}, \bibinfo {author} {\bibfnamefont {A.~M.}\ \bibnamefont {Turner}},
  \bibinfo {author} {\bibfnamefont {A.}~\bibnamefont {Vishwanath}}, \ and\
  \bibinfo {author} {\bibfnamefont {S.~Y.}\ \bibnamefont {Savrasov}},\ }\href
  {\doibase 10.1103/PhysRevB.83.205101} {\bibfield  {journal} {\bibinfo
  {journal} {Phys. Rev. B}\ }\textbf {\bibinfo {volume} {83}},\ \bibinfo
  {pages} {205101} (\bibinfo {year} {2011})}\BibitemShut {NoStop}%
\bibitem [{\citenamefont {Yang}\ \emph {et~al.}(2011)\citenamefont {Yang},
  \citenamefont {Lu},\ and\ \citenamefont {Ran}}]{ying-11}%
  \BibitemOpen
  \bibfield  {author} {\bibinfo {author} {\bibfnamefont {K.-Y.}\ \bibnamefont
  {Yang}}, \bibinfo {author} {\bibfnamefont {Y.-M.}\ \bibnamefont {Lu}}, \ and\
  \bibinfo {author} {\bibfnamefont {Y.}~\bibnamefont {Ran}},\ }\href {\doibase
  10.1103/PhysRevB.84.075129} {\bibfield  {journal} {\bibinfo  {journal} {Phys.
  Rev. B}\ }\textbf {\bibinfo {volume} {84}},\ \bibinfo {pages} {075129}
  (\bibinfo {year} {2011})}\BibitemShut {NoStop}%
\bibitem [{\citenamefont {Delplace}\ \emph {et~al.}(2012)\citenamefont
  {Delplace}, \citenamefont {Li},\ and\ \citenamefont {Carpentier}}]{carp-12}%
  \BibitemOpen
  \bibfield  {author} {\bibinfo {author} {\bibfnamefont {P.}~\bibnamefont
  {Delplace}}, \bibinfo {author} {\bibfnamefont {J.}~\bibnamefont {Li}}, \ and\
  \bibinfo {author} {\bibfnamefont {D.}~\bibnamefont {Carpentier}},\ }\href
  {http://stacks.iop.org/0295-5075/97/i=6/a=67004} {\bibfield  {journal}
  {\bibinfo  {journal} {Europhysics Letters}\ }\textbf {\bibinfo {volume}
  {97}},\ \bibinfo {pages} {67004} (\bibinfo {year} {2012})}\BibitemShut
  {NoStop}%
\bibitem [{\citenamefont {Meng}\ and\ \citenamefont
  {Balents}(2012)}]{balents1}%
  \BibitemOpen
  \bibfield  {author} {\bibinfo {author} {\bibfnamefont {T.}~\bibnamefont
  {Meng}}\ and\ \bibinfo {author} {\bibfnamefont {L.}~\bibnamefont {Balents}},\
  }\href {\doibase 10.1103/PhysRevB.86.054504} {\bibfield  {journal} {\bibinfo
  {journal} {Phys. Rev. B}\ }\textbf {\bibinfo {volume} {86}},\ \bibinfo
  {pages} {054504} (\bibinfo {year} {2012})}\BibitemShut {NoStop}%
\bibitem [{\citenamefont {Burkov}\ and\ \citenamefont
  {Balents}(2011)}]{balents2}%
  \BibitemOpen
  \bibfield  {author} {\bibinfo {author} {\bibfnamefont {A.~A.}\ \bibnamefont
  {Burkov}}\ and\ \bibinfo {author} {\bibfnamefont {L.}~\bibnamefont
  {Balents}},\ }\href {\doibase 10.1103/PhysRevLett.107.127205} {\bibfield
  {journal} {\bibinfo  {journal} {Phys. Rev. Lett.}\ }\textbf {\bibinfo
  {volume} {107}},\ \bibinfo {pages} {127205} (\bibinfo {year}
  {2011})}\BibitemShut {NoStop}%
\bibitem [{\citenamefont {Vafek}\ and\ \citenamefont
  {Vishwanath}(2014)}]{ashvin}%
  \BibitemOpen
  \bibfield  {author} {\bibinfo {author} {\bibfnamefont {O.}~\bibnamefont
  {Vafek}}\ and\ \bibinfo {author} {\bibfnamefont {A.}~\bibnamefont
  {Vishwanath}},\ }\href {\doibase 10.1146/annurev-conmatphys-031113-133841}
  {\bibfield  {journal} {\bibinfo  {journal} {Annual Review of Condensed Matter
  Physics}\ }\textbf {\bibinfo {volume} {5}},\ \bibinfo {pages} {83} (\bibinfo
  {year} {2014})}\BibitemShut {NoStop}%
\bibitem [{\citenamefont {Lababidi}\ and\ \citenamefont {Zhao}(2011)}]{erhai2}%
  \BibitemOpen
  \bibfield  {author} {\bibinfo {author} {\bibfnamefont {M.}~\bibnamefont
  {Lababidi}}\ and\ \bibinfo {author} {\bibfnamefont {E.}~\bibnamefont
  {Zhao}},\ }\href {\doibase 10.1103/PhysRevB.83.184511} {\bibfield  {journal}
  {\bibinfo  {journal} {Phys. Rev. B}\ }\textbf {\bibinfo {volume} {83}},\
  \bibinfo {pages} {184511} (\bibinfo {year} {2011})}\BibitemShut {NoStop}%
\bibitem [{\citenamefont {{Michelsen}}\ and\ \citenamefont
  {{Grein}}(2012)}]{grein}%
  \BibitemOpen
  \bibfield  {author} {\bibinfo {author} {\bibfnamefont {J.}~\bibnamefont
  {{Michelsen}}}\ and\ \bibinfo {author} {\bibfnamefont {R.}~\bibnamefont
  {{Grein}}},\ }\href@noop {} {} (\bibinfo {year} {2012}),\ \Eprint
  {http://arxiv.org/abs/1208.1090} {arXiv:1208.1090 [cond-mat.supr-con]}
  \BibitemShut {NoStop}%
\bibitem [{\citenamefont {Tsutsumi}\ \emph {et~al.}(2011)\citenamefont
  {Tsutsumi}, \citenamefont {Ichioka},\ and\ \citenamefont
  {Machida}}]{PhysRevB.83.094510}%
  \BibitemOpen
  \bibfield  {author} {\bibinfo {author} {\bibfnamefont {Y.}~\bibnamefont
  {Tsutsumi}}, \bibinfo {author} {\bibfnamefont {M.}~\bibnamefont {Ichioka}}, \
  and\ \bibinfo {author} {\bibfnamefont {K.}~\bibnamefont {Machida}},\ }\href
  {\doibase 10.1103/PhysRevB.83.094510} {\bibfield  {journal} {\bibinfo
  {journal} {Phys. Rev. B}\ }\textbf {\bibinfo {volume} {83}},\ \bibinfo
  {pages} {094510} (\bibinfo {year} {2011})}\BibitemShut {NoStop}%
\bibitem [{\citenamefont {Silaev}\ and\ \citenamefont
  {Volovik}(2012)}]{PhysRevB.86.214511}%
  \BibitemOpen
  \bibfield  {author} {\bibinfo {author} {\bibfnamefont {M.~A.}\ \bibnamefont
  {Silaev}}\ and\ \bibinfo {author} {\bibfnamefont {G.~E.}\ \bibnamefont
  {Volovik}},\ }\href {\doibase 10.1103/PhysRevB.86.214511} {\bibfield
  {journal} {\bibinfo  {journal} {Phys. Rev. B}\ }\textbf {\bibinfo {volume}
  {86}},\ \bibinfo {pages} {214511} (\bibinfo {year} {2012})}\BibitemShut
  {NoStop}%
\bibitem [{\citenamefont {Zhao}\ \emph {et~al.}(2010)\citenamefont {Zhao},
  \citenamefont {Zhang},\ and\ \citenamefont {Lababidi}}]{erhai1}%
  \BibitemOpen
  \bibfield  {author} {\bibinfo {author} {\bibfnamefont {E.}~\bibnamefont
  {Zhao}}, \bibinfo {author} {\bibfnamefont {C.}~\bibnamefont {Zhang}}, \ and\
  \bibinfo {author} {\bibfnamefont {M.}~\bibnamefont {Lababidi}},\ }\href
  {\doibase 10.1103/PhysRevB.82.205331} {\bibfield  {journal} {\bibinfo
  {journal} {Phys. Rev. B}\ }\textbf {\bibinfo {volume} {82}},\ \bibinfo
  {pages} {205331} (\bibinfo {year} {2010})}\BibitemShut {NoStop}%
\bibitem [{\citenamefont {Zhang}\ \emph {et~al.}(2009)\citenamefont {Zhang},
  \citenamefont {Liu}, \citenamefont {Qi}, \citenamefont {Dai}, \citenamefont
  {Fang},\ and\ \citenamefont {Zhang}}]{zhang2009}%
  \BibitemOpen
  \bibfield  {author} {\bibinfo {author} {\bibfnamefont {H.}~\bibnamefont
  {Zhang}}, \bibinfo {author} {\bibfnamefont {C.-X.}\ \bibnamefont {Liu}},
  \bibinfo {author} {\bibfnamefont {X.-L.}\ \bibnamefont {Qi}}, \bibinfo
  {author} {\bibfnamefont {X.}~\bibnamefont {Dai}}, \bibinfo {author}
  {\bibfnamefont {Z.}~\bibnamefont {Fang}}, \ and\ \bibinfo {author}
  {\bibfnamefont {S.-C.}\ \bibnamefont {Zhang}},\ }\href {\doibase
  http://dx.doi.org/10.1038/nphys1270} {\bibfield  {journal} {\bibinfo
  {journal} {Nat Phys}\ }\textbf {\bibinfo {volume} {5}},\ \bibinfo {pages}
  {438} (\bibinfo {year} {2009})}\BibitemShut {NoStop}%
\bibitem [{\citenamefont {Chen}\ \emph {et~al.}(2010)\citenamefont {Chen},
  \citenamefont {Chu}, \citenamefont {Analytis}, \citenamefont {Liu},
  \citenamefont {Igarashi}, \citenamefont {Kuo}, \citenamefont {Qi},
  \citenamefont {Mo}, \citenamefont {Moore}, \citenamefont {Lu}, \citenamefont
  {Hashimoto}, \citenamefont {Sasagawa}, \citenamefont {Zhang}, \citenamefont
  {Fisher}, \citenamefont {Hussain},\ and\ \citenamefont
  {Shen}}]{Chen06082010}%
  \BibitemOpen
  \bibfield  {author} {\bibinfo {author} {\bibfnamefont {Y.~L.}\ \bibnamefont
  {Chen}}, \bibinfo {author} {\bibfnamefont {J.-H.}\ \bibnamefont {Chu}},
  \bibinfo {author} {\bibfnamefont {J.~G.}\ \bibnamefont {Analytis}}, \bibinfo
  {author} {\bibfnamefont {Z.~K.}\ \bibnamefont {Liu}}, \bibinfo {author}
  {\bibfnamefont {K.}~\bibnamefont {Igarashi}}, \bibinfo {author}
  {\bibfnamefont {H.-H.}\ \bibnamefont {Kuo}}, \bibinfo {author} {\bibfnamefont
  {X.~L.}\ \bibnamefont {Qi}}, \bibinfo {author} {\bibfnamefont {S.~K.}\
  \bibnamefont {Mo}}, \bibinfo {author} {\bibfnamefont {R.~G.}\ \bibnamefont
  {Moore}}, \bibinfo {author} {\bibfnamefont {D.~H.}\ \bibnamefont {Lu}},
  \bibinfo {author} {\bibfnamefont {M.}~\bibnamefont {Hashimoto}}, \bibinfo
  {author} {\bibfnamefont {T.}~\bibnamefont {Sasagawa}}, \bibinfo {author}
  {\bibfnamefont {S.~C.}\ \bibnamefont {Zhang}}, \bibinfo {author}
  {\bibfnamefont {I.~R.}\ \bibnamefont {Fisher}}, \bibinfo {author}
  {\bibfnamefont {Z.}~\bibnamefont {Hussain}}, \ and\ \bibinfo {author}
  {\bibfnamefont {Z.~X.}\ \bibnamefont {Shen}},\ }\href {\doibase
  10.1126/science.1189924} {\bibfield  {journal} {\bibinfo  {journal}
  {Science}\ }\textbf {\bibinfo {volume} {329}},\ \bibinfo {pages} {659}
  (\bibinfo {year} {2010})}\BibitemShut {NoStop}%
\bibitem [{\citenamefont {Chang}\ \emph {et~al.}(2013)\citenamefont {Chang},
  \citenamefont {Zhang}, \citenamefont {Feng}, \citenamefont {Shen},
  \citenamefont {Zhang}, \citenamefont {Guo}, \citenamefont {Li}, \citenamefont
  {Ou}, \citenamefont {Wei}, \citenamefont {Wang}, \citenamefont {Ji},
  \citenamefont {Feng}, \citenamefont {Ji}, \citenamefont {Chen}, \citenamefont
  {Jia}, \citenamefont {Dai}, \citenamefont {Fang}, \citenamefont {Zhang},
  \citenamefont {He}, \citenamefont {Wang}, \citenamefont {Lu}, \citenamefont
  {Ma},\ and\ \citenamefont {Xue}}]{Chang12042013}%
  \BibitemOpen
  \bibfield  {author} {\bibinfo {author} {\bibfnamefont {C.-Z.}\ \bibnamefont
  {Chang}}, \bibinfo {author} {\bibfnamefont {J.}~\bibnamefont {Zhang}},
  \bibinfo {author} {\bibfnamefont {X.}~\bibnamefont {Feng}}, \bibinfo {author}
  {\bibfnamefont {J.}~\bibnamefont {Shen}}, \bibinfo {author} {\bibfnamefont
  {Z.}~\bibnamefont {Zhang}}, \bibinfo {author} {\bibfnamefont
  {M.}~\bibnamefont {Guo}}, \bibinfo {author} {\bibfnamefont {K.}~\bibnamefont
  {Li}}, \bibinfo {author} {\bibfnamefont {Y.}~\bibnamefont {Ou}}, \bibinfo
  {author} {\bibfnamefont {P.}~\bibnamefont {Wei}}, \bibinfo {author}
  {\bibfnamefont {L.-L.}\ \bibnamefont {Wang}}, \bibinfo {author}
  {\bibfnamefont {Z.-Q.}\ \bibnamefont {Ji}}, \bibinfo {author} {\bibfnamefont
  {Y.}~\bibnamefont {Feng}}, \bibinfo {author} {\bibfnamefont {S.}~\bibnamefont
  {Ji}}, \bibinfo {author} {\bibfnamefont {X.}~\bibnamefont {Chen}}, \bibinfo
  {author} {\bibfnamefont {J.}~\bibnamefont {Jia}}, \bibinfo {author}
  {\bibfnamefont {X.}~\bibnamefont {Dai}}, \bibinfo {author} {\bibfnamefont
  {Z.}~\bibnamefont {Fang}}, \bibinfo {author} {\bibfnamefont {S.-C.}\
  \bibnamefont {Zhang}}, \bibinfo {author} {\bibfnamefont {K.}~\bibnamefont
  {He}}, \bibinfo {author} {\bibfnamefont {Y.}~\bibnamefont {Wang}}, \bibinfo
  {author} {\bibfnamefont {L.}~\bibnamefont {Lu}}, \bibinfo {author}
  {\bibfnamefont {X.-C.}\ \bibnamefont {Ma}}, \ and\ \bibinfo {author}
  {\bibfnamefont {Q.-K.}\ \bibnamefont {Xue}},\ }\href {\doibase
  10.1126/science.1234414} {\bibfield  {journal} {\bibinfo  {journal}
  {Science}\ }\textbf {\bibinfo {volume} {340}},\ \bibinfo {pages} {167}
  (\bibinfo {year} {2013})}\BibitemShut {NoStop}%
\bibitem [{\citenamefont {Zhang}\ \emph {et~al.}(2013)\citenamefont {Zhang},
  \citenamefont {Chang}, \citenamefont {Tang}, \citenamefont {Zhang},
  \citenamefont {Feng}, \citenamefont {Li}, \citenamefont {Wang}, \citenamefont
  {Chen}, \citenamefont {Liu}, \citenamefont {Duan}, \citenamefont {He},
  \citenamefont {Xue}, \citenamefont {Ma},\ and\ \citenamefont
  {Wang}}]{Zhang29032013}%
  \BibitemOpen
  \bibfield  {author} {\bibinfo {author} {\bibfnamefont {J.}~\bibnamefont
  {Zhang}}, \bibinfo {author} {\bibfnamefont {C.-Z.}\ \bibnamefont {Chang}},
  \bibinfo {author} {\bibfnamefont {P.}~\bibnamefont {Tang}}, \bibinfo {author}
  {\bibfnamefont {Z.}~\bibnamefont {Zhang}}, \bibinfo {author} {\bibfnamefont
  {X.}~\bibnamefont {Feng}}, \bibinfo {author} {\bibfnamefont {K.}~\bibnamefont
  {Li}}, \bibinfo {author} {\bibfnamefont {L.-l.}\ \bibnamefont {Wang}},
  \bibinfo {author} {\bibfnamefont {X.}~\bibnamefont {Chen}}, \bibinfo {author}
  {\bibfnamefont {C.}~\bibnamefont {Liu}}, \bibinfo {author} {\bibfnamefont
  {W.}~\bibnamefont {Duan}}, \bibinfo {author} {\bibfnamefont {K.}~\bibnamefont
  {He}}, \bibinfo {author} {\bibfnamefont {Q.-K.}\ \bibnamefont {Xue}},
  \bibinfo {author} {\bibfnamefont {X.}~\bibnamefont {Ma}}, \ and\ \bibinfo
  {author} {\bibfnamefont {Y.}~\bibnamefont {Wang}},\ }\href {\doibase
  10.1126/science.1230905} {\bibfield  {journal} {\bibinfo  {journal}
  {Science}\ }\textbf {\bibinfo {volume} {339}},\ \bibinfo {pages} {1582}
  (\bibinfo {year} {2013})}\BibitemShut {NoStop}%
\bibitem [{\citenamefont {Qi}\ \emph {et~al.}(2008)\citenamefont {Qi},
  \citenamefont {Hughes},\ and\ \citenamefont {Zhang}}]{qi-field}%
  \BibitemOpen
  \bibfield  {author} {\bibinfo {author} {\bibfnamefont {X.-L.}\ \bibnamefont
  {Qi}}, \bibinfo {author} {\bibfnamefont {T.~L.}\ \bibnamefont {Hughes}}, \
  and\ \bibinfo {author} {\bibfnamefont {S.-C.}\ \bibnamefont {Zhang}},\ }\href
  {\doibase 10.1103/PhysRevB.78.195424} {\bibfield  {journal} {\bibinfo
  {journal} {Phys. Rev. B}\ }\textbf {\bibinfo {volume} {78}},\ \bibinfo
  {pages} {195424} (\bibinfo {year} {2008})}\BibitemShut {NoStop}%
\bibitem [{\citenamefont {Zhang}\ \emph {et~al.}(2010)\citenamefont {Zhang},
  \citenamefont {He}, \citenamefont {Chang}, \citenamefont {Song},
  \citenamefont {Wang}, \citenamefont {Chen}, \citenamefont {Jia},
  \citenamefont {Fang}, \citenamefont {Dai}, \citenamefont {Shan} \emph
  {et~al.}}]{zhang2009-2}%
  \BibitemOpen
  \bibfield  {author} {\bibinfo {author} {\bibfnamefont {Y.}~\bibnamefont
  {Zhang}}, \bibinfo {author} {\bibfnamefont {K.}~\bibnamefont {He}}, \bibinfo
  {author} {\bibfnamefont {C.-Z.}\ \bibnamefont {Chang}}, \bibinfo {author}
  {\bibfnamefont {C.-L.}\ \bibnamefont {Song}}, \bibinfo {author}
  {\bibfnamefont {L.-L.}\ \bibnamefont {Wang}}, \bibinfo {author}
  {\bibfnamefont {X.}~\bibnamefont {Chen}}, \bibinfo {author} {\bibfnamefont
  {J.-F.}\ \bibnamefont {Jia}}, \bibinfo {author} {\bibfnamefont
  {Z.}~\bibnamefont {Fang}}, \bibinfo {author} {\bibfnamefont {X.}~\bibnamefont
  {Dai}}, \bibinfo {author} {\bibfnamefont {W.-Y.}\ \bibnamefont {Shan}},
  \emph {et~al.},\ }\href {\doibase http://dx.doi.org/10.1038/nphys1689}
  {\bibfield  {journal} {\bibinfo  {journal} {Nature Physics}\ }\textbf
  {\bibinfo {volume} {6}},\ \bibinfo {pages} {584} (\bibinfo {year}
  {2010})}\BibitemShut {NoStop}%
\bibitem [{\citenamefont {Grein}\ \emph {et~al.}(2012)\citenamefont {Grein},
  \citenamefont {Michelsen},\ and\ \citenamefont {Eschrig}}]{grein2}%
  \BibitemOpen
  \bibfield  {author} {\bibinfo {author} {\bibfnamefont {R.}~\bibnamefont
  {Grein}}, \bibinfo {author} {\bibfnamefont {J.}~\bibnamefont {Michelsen}}, \
  and\ \bibinfo {author} {\bibfnamefont {M.}~\bibnamefont {Eschrig}},\ }\href
  {http://stacks.iop.org/1742-6596/391/i=1/a=012149} {\bibfield  {journal}
  {\bibinfo  {journal} {Journal of Physics: Conference Series}\ }\textbf
  {\bibinfo {volume} {391}},\ \bibinfo {pages} {012149} (\bibinfo {year}
  {2012})}\BibitemShut {NoStop}%
\bibitem [{\citenamefont {Wang}\ \emph {et~al.}(2013)\citenamefont {Wang},
  \citenamefont {Ding}, \citenamefont {Fedorov}, \citenamefont {Yao},
  \citenamefont {Li}, \citenamefont {Lv}, \citenamefont {Zhao}, \citenamefont
  {Zhang}, \citenamefont {Xu}, \citenamefont {Schneeloch}, \citenamefont
  {Zhong}, \citenamefont {Ji}, \citenamefont {Wang}, \citenamefont {He},
  \citenamefont {Ma}, \citenamefont {Gu}, \citenamefont {Yao}, \citenamefont
  {Xue}, \citenamefont {Chen},\ and\ \citenamefont {Zhou}}]{Wang2013}%
  \BibitemOpen
  \bibfield  {author} {\bibinfo {author} {\bibfnamefont {E.}~\bibnamefont
  {Wang}}, \bibinfo {author} {\bibfnamefont {H.}~\bibnamefont {Ding}}, \bibinfo
  {author} {\bibfnamefont {A.~V.}\ \bibnamefont {Fedorov}}, \bibinfo {author}
  {\bibfnamefont {W.}~\bibnamefont {Yao}}, \bibinfo {author} {\bibfnamefont
  {Z.}~\bibnamefont {Li}}, \bibinfo {author} {\bibfnamefont {Y.-F.}\
  \bibnamefont {Lv}}, \bibinfo {author} {\bibfnamefont {K.}~\bibnamefont
  {Zhao}}, \bibinfo {author} {\bibfnamefont {L.-G.}\ \bibnamefont {Zhang}},
  \bibinfo {author} {\bibfnamefont {Z.}~\bibnamefont {Xu}}, \bibinfo {author}
  {\bibfnamefont {J.}~\bibnamefont {Schneeloch}}, \bibinfo {author}
  {\bibfnamefont {R.}~\bibnamefont {Zhong}}, \bibinfo {author} {\bibfnamefont
  {S.-H.}\ \bibnamefont {Ji}}, \bibinfo {author} {\bibfnamefont
  {L.}~\bibnamefont {Wang}}, \bibinfo {author} {\bibfnamefont {K.}~\bibnamefont
  {He}}, \bibinfo {author} {\bibfnamefont {X.}~\bibnamefont {Ma}}, \bibinfo
  {author} {\bibfnamefont {G.}~\bibnamefont {Gu}}, \bibinfo {author}
  {\bibfnamefont {H.}~\bibnamefont {Yao}}, \bibinfo {author} {\bibfnamefont
  {Q.-K.}\ \bibnamefont {Xue}}, \bibinfo {author} {\bibfnamefont
  {X.}~\bibnamefont {Chen}}, \ and\ \bibinfo {author} {\bibfnamefont
  {S.}~\bibnamefont {Zhou}},\ }\href {http://dx.doi.org/10.1038/nphys2744
  10.1038/nphys2744
  http://www.nature.com/nphys/journal/v9/n10/abs/nphys2744.html\#supplementary-information}
  {\bibfield  {journal} {\bibinfo  {journal} {Nat Phys}\ }\textbf {\bibinfo
  {volume} {9}},\ \bibinfo {pages} {621} (\bibinfo {year} {2013})}\BibitemShut
  {NoStop}%
\bibitem [{\citenamefont {Zareapour}\ \emph {et~al.}(2012)\citenamefont
  {Zareapour}, \citenamefont {Hayat}, \citenamefont {Zhao}, \citenamefont
  {Kreshchuk}, \citenamefont {Jain}, \citenamefont {Kwok}, \citenamefont {Lee},
  \citenamefont {Cheong}, \citenamefont {Xu}, \citenamefont {Yang},
  \citenamefont {Gu}, \citenamefont {Jia}, \citenamefont {Cava},\ and\
  \citenamefont {Burch}}]{Zareapour2012}%
  \BibitemOpen
  \bibfield  {author} {\bibinfo {author} {\bibfnamefont {P.}~\bibnamefont
  {Zareapour}}, \bibinfo {author} {\bibfnamefont {A.}~\bibnamefont {Hayat}},
  \bibinfo {author} {\bibfnamefont {S.~Y.~F.}\ \bibnamefont {Zhao}}, \bibinfo
  {author} {\bibfnamefont {M.}~\bibnamefont {Kreshchuk}}, \bibinfo {author}
  {\bibfnamefont {A.}~\bibnamefont {Jain}}, \bibinfo {author} {\bibfnamefont
  {D.~C.}\ \bibnamefont {Kwok}}, \bibinfo {author} {\bibfnamefont
  {N.}~\bibnamefont {Lee}}, \bibinfo {author} {\bibfnamefont {S.-W.}\
  \bibnamefont {Cheong}}, \bibinfo {author} {\bibfnamefont {Z.}~\bibnamefont
  {Xu}}, \bibinfo {author} {\bibfnamefont {A.}~\bibnamefont {Yang}}, \bibinfo
  {author} {\bibfnamefont {G.~D.}\ \bibnamefont {Gu}}, \bibinfo {author}
  {\bibfnamefont {S.}~\bibnamefont {Jia}}, \bibinfo {author} {\bibfnamefont
  {R.~J.}\ \bibnamefont {Cava}}, \ and\ \bibinfo {author} {\bibfnamefont
  {K.~S.}\ \bibnamefont {Burch}},\ }\href {http://dx.doi.org/10.1038/ncomms2042
  http://www.nature.com/ncomms/journal/v3/n9/suppinfo/ncomms2042\_S1.html}
  {\bibfield  {journal} {\bibinfo  {journal} {Nat Commun}\ }\textbf {\bibinfo
  {volume} {3}},\ \bibinfo {pages} {1056} (\bibinfo {year} {2012})}\BibitemShut
  {NoStop}%
\bibitem [{\citenamefont {Wang}\ \emph {et~al.}(2012)\citenamefont {Wang},
  \citenamefont {Liu}, \citenamefont {Xu}, \citenamefont {Yang}, \citenamefont
  {Miao}, \citenamefont {Yao}, \citenamefont {Gao}, \citenamefont {Shen},
  \citenamefont {Ma}, \citenamefont {Chen}, \citenamefont {Xu}, \citenamefont
  {Liu}, \citenamefont {Zhang}, \citenamefont {Qian}, \citenamefont {Jia},\
  and\ \citenamefont {Xue}}]{Wang06042012}%
  \BibitemOpen
  \bibfield  {author} {\bibinfo {author} {\bibfnamefont {M.-X.}\ \bibnamefont
  {Wang}}, \bibinfo {author} {\bibfnamefont {C.}~\bibnamefont {Liu}}, \bibinfo
  {author} {\bibfnamefont {J.-P.}\ \bibnamefont {Xu}}, \bibinfo {author}
  {\bibfnamefont {F.}~\bibnamefont {Yang}}, \bibinfo {author} {\bibfnamefont
  {L.}~\bibnamefont {Miao}}, \bibinfo {author} {\bibfnamefont {M.-Y.}\
  \bibnamefont {Yao}}, \bibinfo {author} {\bibfnamefont {C.~L.}\ \bibnamefont
  {Gao}}, \bibinfo {author} {\bibfnamefont {C.}~\bibnamefont {Shen}}, \bibinfo
  {author} {\bibfnamefont {X.}~\bibnamefont {Ma}}, \bibinfo {author}
  {\bibfnamefont {X.}~\bibnamefont {Chen}}, \bibinfo {author} {\bibfnamefont
  {Z.-A.}\ \bibnamefont {Xu}}, \bibinfo {author} {\bibfnamefont
  {Y.}~\bibnamefont {Liu}}, \bibinfo {author} {\bibfnamefont {S.-C.}\
  \bibnamefont {Zhang}}, \bibinfo {author} {\bibfnamefont {D.}~\bibnamefont
  {Qian}}, \bibinfo {author} {\bibfnamefont {J.-F.}\ \bibnamefont {Jia}}, \
  and\ \bibinfo {author} {\bibfnamefont {Q.-K.}\ \bibnamefont {Xue}},\ }\href
  {\doibase 10.1126/science.1216466} {\bibfield  {journal} {\bibinfo  {journal}
  {Science}\ }\textbf {\bibinfo {volume} {336}},\ \bibinfo {pages} {52}
  (\bibinfo {year} {2012})}\BibitemShut {NoStop}%
\bibitem [{\citenamefont {Xu}\ \emph {et~al.}(2012)\citenamefont {Xu},
  \citenamefont {Neupane}, \citenamefont {Liu}, \citenamefont {Zhang},
  \citenamefont {Richardella}, \citenamefont {{Andrew Wray}}, \citenamefont
  {Alidoust}, \citenamefont {Leandersson}, \citenamefont {Balasubramanian},
  \citenamefont {Sanchez-Barriga}, \citenamefont {Rader}, \citenamefont
  {Landolt}, \citenamefont {Slomski}, \citenamefont {{Hugo Dil}}, \citenamefont
  {Osterwalder}, \citenamefont {Chang}, \citenamefont {Jeng}, \citenamefont
  {Lin}, \citenamefont {Bansil}, \citenamefont {Samarth},\ and\ \citenamefont
  {{Zahid Hasan}}}]{Xu2012}%
  \BibitemOpen
  \bibfield  {author} {\bibinfo {author} {\bibfnamefont {S.-Y.}\ \bibnamefont
  {Xu}}, \bibinfo {author} {\bibfnamefont {M.}~\bibnamefont {Neupane}},
  \bibinfo {author} {\bibfnamefont {C.}~\bibnamefont {Liu}}, \bibinfo {author}
  {\bibfnamefont {D.}~\bibnamefont {Zhang}}, \bibinfo {author} {\bibfnamefont
  {A.}~\bibnamefont {Richardella}}, \bibinfo {author} {\bibfnamefont
  {L.}~\bibnamefont {{Andrew Wray}}}, \bibinfo {author} {\bibfnamefont
  {N.}~\bibnamefont {Alidoust}}, \bibinfo {author} {\bibfnamefont
  {M.}~\bibnamefont {Leandersson}}, \bibinfo {author} {\bibfnamefont
  {T.}~\bibnamefont {Balasubramanian}}, \bibinfo {author} {\bibfnamefont
  {J.}~\bibnamefont {Sanchez-Barriga}}, \bibinfo {author} {\bibfnamefont
  {O.}~\bibnamefont {Rader}}, \bibinfo {author} {\bibfnamefont
  {G.}~\bibnamefont {Landolt}}, \bibinfo {author} {\bibfnamefont
  {B.}~\bibnamefont {Slomski}}, \bibinfo {author} {\bibfnamefont
  {J.}~\bibnamefont {{Hugo Dil}}}, \bibinfo {author} {\bibfnamefont
  {J.}~\bibnamefont {Osterwalder}}, \bibinfo {author} {\bibfnamefont {T.-R.}\
  \bibnamefont {Chang}}, \bibinfo {author} {\bibfnamefont {H.-T.}\ \bibnamefont
  {Jeng}}, \bibinfo {author} {\bibfnamefont {H.}~\bibnamefont {Lin}}, \bibinfo
  {author} {\bibfnamefont {A.}~\bibnamefont {Bansil}}, \bibinfo {author}
  {\bibfnamefont {N.}~\bibnamefont {Samarth}}, \ and\ \bibinfo {author}
  {\bibfnamefont {M.}~\bibnamefont {{Zahid Hasan}}},\ }\href
  {http://dx.doi.org/10.1038/nphys2351
  http://www.nature.com/nphys/journal/v8/n8/abs/nphys2351.html\#supplementary-information}
  {\bibfield  {journal} {\bibinfo  {journal} {Nat Phys}\ }\textbf {\bibinfo
  {volume} {8}},\ \bibinfo {pages} {616} (\bibinfo {year} {2012})}\BibitemShut
  {NoStop}%
\bibitem [{\citenamefont {Nakayama}\ \emph {et~al.}(2012)\citenamefont
  {Nakayama}, \citenamefont {Eto}, \citenamefont {Tanaka}, \citenamefont
  {Sato}, \citenamefont {Souma}, \citenamefont {Takahashi}, \citenamefont
  {Segawa},\ and\ \citenamefont {Ando}}]{PhysRevLett.109.236804}%
  \BibitemOpen
  \bibfield  {author} {\bibinfo {author} {\bibfnamefont {K.}~\bibnamefont
  {Nakayama}}, \bibinfo {author} {\bibfnamefont {K.}~\bibnamefont {Eto}},
  \bibinfo {author} {\bibfnamefont {Y.}~\bibnamefont {Tanaka}}, \bibinfo
  {author} {\bibfnamefont {T.}~\bibnamefont {Sato}}, \bibinfo {author}
  {\bibfnamefont {S.}~\bibnamefont {Souma}}, \bibinfo {author} {\bibfnamefont
  {T.}~\bibnamefont {Takahashi}}, \bibinfo {author} {\bibfnamefont
  {K.}~\bibnamefont {Segawa}}, \ and\ \bibinfo {author} {\bibfnamefont
  {Y.}~\bibnamefont {Ando}},\ }\href {\doibase 10.1103/PhysRevLett.109.236804}
  {\bibfield  {journal} {\bibinfo  {journal} {Phys. Rev. Lett.}\ }\textbf
  {\bibinfo {volume} {109}},\ \bibinfo {pages} {236804} (\bibinfo {year}
  {2012})}\BibitemShut {NoStop}%
\bibitem [{\citenamefont {Xu}\ \emph {et~al.}(2014)\citenamefont {Xu},
  \citenamefont {Liu}, \citenamefont {Wang}, \citenamefont {Ge}, \citenamefont
  {Liu}, \citenamefont {Yang}, \citenamefont {Chen}, \citenamefont {Liu},
  \citenamefont {Xu}, \citenamefont {Gao}, \citenamefont {Qian}, \citenamefont
  {Zhang},\ and\ \citenamefont {Jia}}]{jia-13}%
  \BibitemOpen
  \bibfield  {author} {\bibinfo {author} {\bibfnamefont {J.-P.}\ \bibnamefont
  {Xu}}, \bibinfo {author} {\bibfnamefont {C.}~\bibnamefont {Liu}}, \bibinfo
  {author} {\bibfnamefont {M.-X.}\ \bibnamefont {Wang}}, \bibinfo {author}
  {\bibfnamefont {J.}~\bibnamefont {Ge}}, \bibinfo {author} {\bibfnamefont
  {Z.-L.}\ \bibnamefont {Liu}}, \bibinfo {author} {\bibfnamefont
  {X.}~\bibnamefont {Yang}}, \bibinfo {author} {\bibfnamefont {Y.}~\bibnamefont
  {Chen}}, \bibinfo {author} {\bibfnamefont {Y.}~\bibnamefont {Liu}}, \bibinfo
  {author} {\bibfnamefont {Z.-A.}\ \bibnamefont {Xu}}, \bibinfo {author}
  {\bibfnamefont {C.-L.}\ \bibnamefont {Gao}}, \bibinfo {author} {\bibfnamefont
  {D.}~\bibnamefont {Qian}}, \bibinfo {author} {\bibfnamefont {F.-C.}\
  \bibnamefont {Zhang}}, \ and\ \bibinfo {author} {\bibfnamefont {J.-F.}\
  \bibnamefont {Jia}},\ }\href {\doibase 10.1103/PhysRevLett.112.217001}
  {\bibfield  {journal} {\bibinfo  {journal} {Phys. Rev. Lett.}\ }\textbf
  {\bibinfo {volume} {112}},\ \bibinfo {pages} {217001} (\bibinfo {year}
  {2014})}\BibitemShut {NoStop}%
\bibitem [{\citenamefont {Yan}\ \emph {et~al.}(2013)\citenamefont {Yan},
  \citenamefont {Jansen},\ and\ \citenamefont {Felser}}]{Yan2013}%
  \BibitemOpen
  \bibfield  {author} {\bibinfo {author} {\bibfnamefont {B.}~\bibnamefont
  {Yan}}, \bibinfo {author} {\bibfnamefont {M.}~\bibnamefont {Jansen}}, \ and\
  \bibinfo {author} {\bibfnamefont {C.}~\bibnamefont {Felser}},\ }\href
  {http://dx.doi.org/10.1038/nphys2762 10.1038/nphys2762
  http://www.nature.com/nphys/journal/v9/n11/abs/nphys2762.html\#supplementary-information}
  {\bibfield  {journal} {\bibinfo  {journal} {Nat Phys}\ }\textbf {\bibinfo
  {volume} {9}},\ \bibinfo {pages} {709} (\bibinfo {year} {2013})}\BibitemShut
  {NoStop}%
\end{thebibliography}%

\end{document}